\documentclass[12pt,a4paper]{article}

\usepackage{graphicx} 
\usepackage{latexsym}	
\usepackage{amsmath}
\usepackage{amsthm}
\usepackage{amssymb}
\usepackage{amsmath}
\usepackage{amsthm}

\usepackage{graphicx}
\def\ba{\begin{eqnarray}}
\def\ea{\end{eqnarray}}

\def\ba{\begin{eqnarray}}
\def\ea{\end{eqnarray}}

\def\be{\begin{equation}}
\def\ee{\end{equation}}

\theoremstyle{plain}

\begin{document}
\title{Gauge preservation in renormalization for Yang-Mills and gravity theories}

\author{Osvaldo Santillán$^{1}$ and Alejandro Morano$^2$
\thanks{Electronic addresses:  firenzecita@hotmail.com, alejandro.d.morano@gmail.com}\\
\textit{\small{$^1$Instituto de Matem\'atica Luis Santaló (IMAS) and CONICET,}}\\
\textit{\small{Ciudad Universitaria, 1428 Buenos Aires, Argentina.}}\\
\textit{\small{$^2$Instituto de Física de Buenos Aires (IFIBA) }}\\
\textit{\small{and  Departamento de Física UBA}}\\
\textit{\small{Ciudad Universitaria, 1428 Buenos Aires, Argentina.}}}
\date{}

\maketitle

\begin{abstract}
In the present work, multiplicative renormalization \cite{dixon} for Yang-Mills theories is reviewed. While this subject is not new, it is suggested that a clear understanding of these methods leads to a systematic way for interpreting the counter terms needed for non multiplicative renormalization of quantum gravity, for models such as the ones introduced in \cite{dewitt}-\cite{stelle2}. These models are renormalizable, however they contain apparent instabilities leading to possible unitarity loss, an earlier analog is \cite{pais}. The systematic procedure  discussed along the text may be interesting, especially in the modern context, since there are efforts
for avoiding with those apparent instabilities by employing variants of the standard quantization methods \cite{mannheim1}-\cite{salve}.
\end{abstract}
\tableofcontents
\section{Introduction}
The discovery of the renormalizability  of Yang-Mills theories \cite{thooft}-\cite{ash} marked a new era in QFT and particle physics and, in particular, led to the development of the Standard Model. There are  several aspects about confining theories which are still not understood. The consensus is that Yang-Mills theories are correct and that the methods still available are not suited for studying their predictions in full detail. 

Gravity theories on the other hand are more elusive, and the dream of constructing a renormalizable and unitary model of gravitation eluded success for years. There are models  close to achieve several features that a consistent gravity model has to have, however, there is always a missing point which invalidate the model or request further improvements. Quadratic gravity models are examples, these scenarios posses fourth order equations of motion and were considered as a plausible quantum gravity candidates in \cite{dewitt}-\cite{dewitt3}. Their properties were studied further by Stelle in \cite{stelle1}-\cite{stelle2}, in the context of modern QFT.  These classic references, in particular \cite{stelle2} consider the path integral quantization of theories, the problem of gauge fixing and the solution of the Slavnov-Taylor identities for these scenarios. 

The authors of  \cite{dewitt3}-\cite{stelle2} recognize however, that these theories apparently posses a ghost in the physical spectrum. This suggest instabilities and unitary loss \cite{pais}. The point is  propagator of such four order theory will be given in a Pauli Villars form as
$$
D\sim \frac{1}{k^2}-\frac{1}{k^2+M^2}.
$$
For  large values of $k$ this propagator will go as $M^2 k^{-4}$. In QFT
without gravity, the mass $M$ is interpreted as a cutoff, which tends $M\to\infty$ at the end of calculations. In the gravity approach, this cutoff  is rendered  finite, and the minus sign in the last propagator expression 
suggests the presence of a state with negative kinetic energy, which may indicate a potential instability. 

Even taking into account this apparent drawback, it should be mentioned that this type of view about instabilities have been discussed in some modern references. An incomplete list is  \cite{mannheim1}-\cite{donogue}. These references suggest that the problem of the apparent instabilities may be solved by employing variants of the quantization method, some of them inspired in earlier works of Pauli and Dirac. This revivals the interest in those quadratic models since, if the ghost problem is overcome, they will become a very promising candidate for a quantum gravity model. 

The renewed interest in the subject is  one of the motivations for the present work. The purpose is to describe how the results of \cite{stelle1}-\cite{stelle2} ensure that BRST invariance \cite{brst} is preserved order by order in the calculation of the quantum effective action. At first sight, this topic has to be already understood. On the other hand, the authors feel that  still is worthy to write about the topic  for the following reasons. The Slavnov-Taylor identities are consequences of the BRST invariance of the model. However, the Slavnov-Taylor identities alone do not insure that order by order the effective quantum action is invariant under diffeomorphisms. In Yang-Mills theories, this is established by multiplicative renormalization. The renormalization of quadratic gravity however, is not multiplicative. The reference \cite{stelle2} derives the solution of the Slavnov-Taylor identities and establish succinctly that the non gauge terms can be canceled by clever field redefinitions. While we fully agree with the results of this reference, the intention is to give an interpretation of those field redefinitions in a more systematic way. 

In fact,  although there exist some literature reviewing this subject, we believe that the subject is nowhere fully discussed. There are detailed references about quantum gravity available such as \cite{buchbinder}-\cite{hamber} which covers several important aspects of those theories in great detail. However, this particular topic is only covered partially. In addition, several references about quantum field theory cover aspects of multiplicative renormalization  \cite{weinberg}-\cite{bilal}  with a formalism which enlarge the BRST symmetry with the action of auxiliary fields. This procedure has several advantages and it is useful for studying several types of gauges. However, this is not exactly the formalism for which the gravity results of \cite{stelle2} are formulated. There are other references such as \cite{itzykson}-\cite{ryder} which covers the subject partially, although they contain clear expositions about the solutions of the Slavnov-Taylor identities. The purpose is to fill this small gap as much as possible by going through the calculations establishing multiplicative renormalization and the approximate order by order preservation of BRST invariance by the renormalization procedure. The advantage is that once these details are understood for Yang-Mills theories, even though gravity  is not renormalized multiplicatively, there are analogies that lead directly  to the field redefinition which ensure order by order approximate BRST invariance, found in \cite{stelle2}. We consider this systematic therefore of interest, especially in the present context, as it allows an interpretation of the otherwise obscure counter terms that appear in gravity.

The reader will probably notice that the present exposition lacks details in some aspects and is pretty detailed in some others. The details that are omitted can be found in \cite{weinberg}-\cite{ryder} or \cite{buchbinder}-\cite{hamber}, and it is not the intention to compete with  those clear expositions. The detailed parts of the present text correspond to situations which in our opinion are not discussed in the available literature and are potentially useful. The idea is that the reader may complement the literature with the present details in order to get some new insights. It may be said that the text contains the explanations that we would have liked to have when learning the subject.  The reward of the effort is that an interpretation about the counter terms in the gravity scenario appears that, at the best of the authors knowledge, is not particularly described in the available literature. Recent modern literature dealing with topics of interest such as the background field method or 
renormalization group may be found in \cite{shapi1}-\cite{shapi5}.

The present work is briefly organized as follows. Section 2 discusses the Slavnov-Taylor identities in the context of Yang-Mills theories, and contains detailed explanations about how multiplicative renormalization is achieved and how it insures BRST invariance at a given order in $\hbar$. Section 3 contains the arguments about the renormalizability of quadratic gravity, and the explicit solutions of the Slavnov-Taylor identities for these models. The solution is interpreted by keeping some analogy with multiplicative renormalization. Even in the gravity case the renormalization is not multiplicative, it is shown that BRST invariance is preserved order by order by the addition of non standard coupling to the initial action, which still respect BRST  symmetry.
The last section contains some possible applications.

\pagebreak

\section{The renormalization program for Yang-Mills theories}
\subsection{Preliminars about Yang-Mills theories}
The topic to be studied here is a pure Yang-Mills theory with a gauge field given by $A=A_\mu^a T_a$
where $T_a$ are a representation of a  Lie algebra corresponding  to some Lie group, denominated the gauge group, not necessarily abelian. The elements $T_a$ satisfy commutation relations of the form $[T_a, T_b]= ic_{abc} T_c$ with $c_{abc}$ the structure constants corresponding to the given basis $T_a$. By defining the strength tensor
$$
 F_{\mu\nu}^a=\partial_\mu A^a_\nu-\partial_\nu A^a_\mu+gc_{abc}A^b_\mu A_\nu^c,
 $$
the Yang-Mills lagrangian is given by
$$
L_y=-\frac{1}{4} F_{\mu\nu}^a F^{\mu\nu a}.
$$
This lagrangian  is invariant under the gauge transformation  
\begin{equation}\label{gauge}
A\to A^G=G A G^{-1}-\frac{i}{g}(\partial_\mu G) G^{-1},
\end{equation}
 with $G$ an element of the gauge group
$$
G=e^{-i T_a \theta^a}, \qquad a=1,., \text{dim}\;G,
$$
with $\theta^a(x)$ space time dependent coordinates parameterizing the gauge group manifold. For a non abelian algebra, that is, for $c_{abc}\neq 0$, the gauge field interacts with itself due to the term $gc_{abc}A^b_\mu A_\nu^c$ which, as it is known, generates cubic and quartic terms interaction vertices. These vertices however should be found only after fixing the gauge.

At infinitesimal level the gauge transformation \eqref{gauge} may be expressed as
\begin{equation}\label{variando}
 \delta A_\nu^a=\frac{1}{g}D_\mu^{ab} \theta^b,\qquad D_\mu^{ab}=\delta^{ab} \partial_\mu-g c_{abc}A_\mu^c,
\end{equation}
the operator $D_\mu^{ab}$ is known as the covariant derivative corresponding to the gauge field. In these terms, it is possible to obtain a necessary condition for a given functional $s$ to be gauge invariant. Assume that a functional
$$
s=\int d^4x L(A_\mu^\alpha, \partial_\nu A_\mu^\alpha), 
$$
is given with $ L(A_\mu^\alpha, \partial_\nu A_\mu^\alpha)$ not necessarily identified with the Yang-Mills lagrangian. If this functional $s$ is gauge invariant,  then under a gauge variation \eqref{variando} of the fields
$$
\int d^4x \frac{\delta s}{\delta A_\mu^a} D_\mu^{ab}\theta^b=0.
$$
Here $\frac{\delta s}{\delta A_\mu^a}$ is the Euler-Lagrange expression, and the classical equations of motion are identified the vanishing of this expression. After an integration by parts of the derivatives inside $D_\mu^{ab}$ in the last formula, it is found that
$$
\int d^4x D_\mu^{ab}\bigg(\frac{\delta s}{\delta A_\mu^a}\bigg) \theta^b=0.
$$
Since the last condition has to be valid for every $\theta^b$, it follows that a gauge invariant functional must satisfy
\begin{equation}\label{gubki}
 D_\mu^{ab}\bigg(\frac{\delta s}{\delta A_\mu^a}\bigg)=0.
\end{equation}
The  conclusion is that the  equations of motion for any gauge invariant action  $s$ are always annihilated by the action of the covariant derivative $D_\mu^{ab}$, and it is not difficult to see that the converse is also true. 

There exists a tower of possible gauge invariant actions $s$ that can be constructed in terms of the gauge field $A_\mu^a$. The particularity of the Yang-Mills choice is that the lagrangian  has mass dimension equal to four and it leads to a  renormalizable theory. Roughly speaking, renormalizable  means that the number of divergent diagrams that appear when calculating a given Green function  can be cured by use of a finite number of counter-terms in the lagrangian. If the mass dimension is larger than four, then a mass dependent coupling constant appears and renormalizabiltiy may be spoiled, for instance, as in the old Fermi theory with $G_f\sim M_W^{-2}$. 

Even though  Yang-Mills models are renormalizable, one of the key points was to understand if the counter terms involved preserve order by order  a generalization of  gauge invariance, named BRST invariance \cite{brst}, which holds when the action is gauge fixed. An important tool for studying this problem are the Slavnov-Taylor identities \cite{slavnov}-\cite{taylor}, whose consequences for gauge and gravity theories will be described in the next chapters.

\subsection{Peculiarities about gauge fixing and quantization}
One of the widely employed methods for quantizing gauge theories is the Feynman path integral
\begin{equation}\label{feynman}
Z(J)=\int DA_\mu e^{i S_y+i\int d^4x J^\mu A^a_\mu}.
\end{equation}
This integral sum a fixed configuration $A^{a f}_\mu(x)$ and all its gauge transformed representatives $A^{a f G}_\mu(x)$.  However, the integral one would like to calculate is 
\begin{equation}\label{henman}
Z_g(J)=\int [DA_\mu] e^{i [S_y+i\int d^4x J^\mu A^a_\mu]},
\end{equation}
with the bracket $[]$ indicating one representative of a full gauge transformed configuration. The resulting measure $[DA_\mu]$, as it is well known and will be described below, contains non trivial factors. As there is a plethora of such transformed representatives, their contribution leads to a divergent value of $Z(J)$.  The cure for this problem is to impose a gauge fixing, which is a procedure allowing to express the divergent part of $Z(J)$ as an  overall divergent factor that can be discarded. Explicitly, it is expected that
$$
Z(J)=Vol(G)Z_g(J).
$$
Leaving aside issues such as the Gribov ambiguities \cite{gribov}, which may obstruct the standard gauge fixing procedure, in the following, linear gauges of the form $F^a(A^b)=0$ with
$$
F_a(A_\mu^b)=\phi^\mu_{ab} A_{\mu b},
$$
will be considered. The  symbol $\phi_{ab}^\mu$  denotes an operator which is  linear in space time derivatives. For instance, the Feynman gauge corresponds to the choice $\phi_{ab}^\mu=\partial_\mu \delta_{ab}$. This gauge choice simplifies several aspects of quantization. It is clear that for these linear gauges the following relation holds
\begin{equation}\label{facilongo}
\frac{\delta F_a(A^c_\nu)}{\delta A^b_\mu}=\phi_{ab}^\mu,
\end{equation}
as $\phi_{ab}^\mu$ does not depend on $A^a_\mu$.

Given a gauge fixing  $F_a(A_\mu^b)$, there are gauge transformations that leave the value of $F^a(A^b_\mu)$ unchanged, and there are others that do not.  This depends on the choice of $\phi_{ab}^\nu$.  The most familiar example takes place in Maxwell theory with the Coulomb gauge. In this case, any gauge transformations changes the value of $F^a(A^b_\mu)$. This is not the case for the Lorenz gauge, where there exist residual gauge transformation leaving $F^a(A^b_\mu)$ fixed.  In other words the Coulomb gauge fixes the potential vector completely while the Lorenz gauge does not.  

In a non abelian theory, a gauge transformation \eqref{gauge} is induced by a group element, $G$ corresponding to parameters $\theta^a$ introduced in that formula. Under a small gauge transformation it generally given as
\begin{equation}\label{ryder}
F_a(A^{bG}(x))=F_a(A^b(x))+\int M_{ac}(x, y)\theta^c(y) d^4y,
\end{equation}
where $M_{ac}(x, y)$ some matrix, whose functional form depends on the choice of $F^a$. Some of these choices of $\theta^c$ will give a vanishing contribution,  $$\int M_{ac}(x, y)\theta^c(y) d^4y=0,$$  and these correspond to transformations that leave the gauge surface invariant. In a general situation, the additional contribution is not zero.  In order to give an example consider again the abelian Maxwell theory with the Lorenz gauge $F(A^\mu)=\partial_\mu A^\mu=0$.  This corresponds to the simple choice $\phi_\mu=\partial_\mu$.  Under a gauge transformation $A_\mu\to A_\mu^\alpha=A_\mu+\partial_\mu \theta(x)$ it is direct to find that
$$
F(A^{\mu(\alpha)})-F(A^\mu)=\partial_\mu \partial^\mu \theta(x),
$$
from where
$$
\frac{\delta F}{\delta \theta(y)}=\partial_\mu \partial^\mu \delta^4(x-y),
$$
and therefore
\begin{equation}\label{aimitar2}
F(A^{\mu(\alpha)})-F(A^\mu)=\int \partial_\mu \partial^\mu \delta^4(x-y) \theta(y) d^4y=\partial_\mu \partial^\mu \theta(x).
\end{equation}
From the last expression it is calculated that
\begin{equation}\label{tatu}
M(x,y)=\partial_\mu \partial^\mu \delta^4(x-y).
\end{equation}
In this gauge it is clear that  the functions $\theta(x)$ satisfying the second order equation $\partial_\mu \partial^\mu \theta(x)=0$ do not change the value of $F(A^\mu)=\partial_\mu A^\mu$, this is seen directly from \eqref{aimitar2}. This reflects the comment given above, the Lorenz gauge does not completely fix the potential. Instead the more restrictive Coulomb gauge does fix it.

The advantages of the Coulomb gauge in the abelian theory do not hold for non abelian ones. Already in the $SU(2)$ model the Coulomb gauge does not fix completely the gauge potential and, to the best of the authors knowledge, there is no gauge known such that this complete fixing is achieved. On the other hand, this problem known as Gribov ambiguity \cite{gribov} is expected to complicate non perturbative aspects of the theory. This important issue will not be considered further here, by assuming (without full justification) that the Gribov ambiguity does not affect the perturbative calculations.

The use of an operator $M(x,y)$  in \eqref{ryder} depending on two space time variables $x$ and $y$  has some advantages in path integral quantization. If one prefers an operator depending on one space time coordinate, then by expressing formula \eqref{aimitar2} as
$$
\delta F(A^\mu)=\phi_\mu\delta A^\mu(x)=\phi_\mu \partial^\mu\theta(x),
$$
and by introducing  the operator  $M(x)=\phi_\mu\partial_\mu$ it follows that $\delta F(A^\mu)=M(x)\theta(x)$. In a non abelian theory the last formula is generalized by replacing the derivative $\partial_\mu$ with the covariant one $D_{\mu}^{ab}$, the result is
\begin{equation}\label{noabel}
\delta F^a(A_\mu^b)=\phi^\mu_{ac} D_{\mu}^{cb}\theta_b(x),
\end{equation}
where explicitly $D_{\mu}^{ab}=\partial_\mu\delta^{ab}-g c_{abc}A_{\mu}^c$. With the further identification 
\begin{equation}\label{identificatiion}
M(x)=\phi^\mu_{ac} D_{\mu}^{cb},
\end{equation}
the last equation reads
\begin{equation}\label{identification2}
    \delta F^a=M_{ab}(x) \theta^b.
\end{equation}
The formulas \eqref{identificatiion} and \eqref{identification2} are important role in the developments of the  next sections.

The gauge-fixing formulas outlined above play a crucial role in understanding an essential tool in Feynman path integral quantization: the functional Dirac delta distributions. When dealing with gauge fixing there appear factors such as  $\delta[F_a(A^{bG}(x))-F_a(A^b(x))]$  inside the path integral, which define the chosen gauge surface.  The standard Fourier representation of a n-dimensional  Dirac delta $$\delta^n(x_1, .., x_n)=\int  e^{i<p, x>} dp_1..dp_n,\qquad <p, x>=p_i x^i,$$ can be generalized to the functional context to formulas of the following type
$$
\delta[F_a(A^{bG}(x))-F_a(A^b(x))]=\delta(\int M_{ac}(x, y)\theta^c(y) d^4y)=\int Dp^a
e^{i\int p^a (x)M_{ac}(x, y)\theta^c(y) d^4y dx^4}.
$$
Furthermore, from the functional generalization of the Gaussian multi dimensional integrals $$\int e^{-x^a M_{ab} y^b}dx^a dy^b=\text{Det}^{-1}(M_{ab}),$$ to the continuum  case, it follows that the integration over $\theta(x)$ of the last formula, leads to
\begin{equation}\label{determinante-}
\int  D\theta^c \delta[F_a(A^{b\theta^c}(x))-F_a(A^b(x))]=\int Dp^a D\theta^c
e^{i\int p^a (x)M_{ac}(x, y)\theta^c(y) d^4y dx^4}=\text{Det}^{-1}(iM_{ab}(x,y)).
\end{equation}
The inverse of the determinant of the operator $M_{ab}$ can be interpreted as the product of its inverse eigenvalues $\lambda_i^{-1}$, when they exist and are non zero. 

If instead of real variables $x^a$ and $y^a$, Grassmann variables $\eta^a$ and $\overline{\eta}^a$, that is, anti-comunting variables 
$$
\eta^a\eta^b=-\eta^b\eta^a,\qquad  \overline{\eta}^a\overline{\eta}^b=-\overline{\eta}^b\overline{\eta}^a, \qquad \overline{\eta}^a\eta^b=-\eta^b\overline{\eta}^a.
$$ 
are employed, then there exist formulas such as $$\int e^{i\eta^a M_{ab} \overline{\eta}^b}d\eta^a d\overline{\eta}^b=\text{Det}(iM_{ab}).$$ By comparing the last three formulas, it is concluded that the change from real to Grassmann variables converts the determinant of a matrix into its inverse.  The proof is not difficult, since the exponent of the last integral becomes an expression of first order in the Grassman variables  due to the nilpotency conditions $\eta^a\eta^a=0$.  A Taylor expansion of the integrand leads to the desired result. With the previous remarks in mind, the equation \eqref{determinante-} can be read in reversed order, in order to conclude that
\begin{equation}\label{determinante}
\text{Det}(iM_{ab}(x,y))=\int D\overline{\eta}^a D\eta^c
e^{-i\int \overline{\eta}^a (x)M_{ac}(x, y)\eta^c(y) d^4y dx^4}.
\end{equation}
Following the standard denomination, the variables $\eta^a$ and $\overline{\eta}^a$ will be referred as Faddeev-Popov ghost and anti-ghost respectively.  Note that one of the integrals in the exponent, for example the $y$ integration, may be eliminated due to the Dirac delta factor $\delta^4(x-y)$ that usually appears in the expression for $M_{ab}(x, y)$. 

Given these determinantal formulas, the strategy of the gauge fixing procedure is to convert the divergence present in \eqref{feynman} into an overall normalization factor, also divergent, which can be factored out. By assuming that a gauge fixing operator $F^a(A_\mu^b)$  that fixes the gauge completely has been found one may insert a unit  inside $Z(J)$ represented in the form
$$
1=\Delta_{FP}(A^{bG}(x))\int DG  \delta[F_a(A^{bG}(x))-F_a(A^b(x))],
$$
where the Faddev-Popov factor $\Delta_{FP}(A^b)$ has been introduced.  The Faddev-Popov determinant is whatever is needed for making the last expression equal to one , this ensures that the insertion of the last factor inside $Z(J)$ will not change its value. The Dirac delta dictates the choice of gauge, as depends on $F^a(A_\mu^b)$ explicitly.   On the other hand, the formulas \eqref{determinante-} and \eqref{determinante} imply that this factor may be represented in terms of ghost and anti-ghost as
\begin{equation}\label{popin}
\Delta_{FP}=\text{Det}(M_{ab}(x,y))=\int D\overline{\eta}^a D\eta^c
e^{-i\int \overline{\eta}^a (x)M_{ac}(x, y)\eta^c(y) d^4y dx^4}.
\end{equation}
The Faddeev-Popov determinant in fact is gauge invariant, $\Delta_{FP}(A^{bG}(x))=\Delta_{FP}(A^{b}(x))$, the proof relies in the use of the Haar measure for gauge groups. Proofs can be found for instance,  in the pages 20-22 of \cite{weinberg} or pages 299-301 in \cite{kaku}.  In these terms, the functional \eqref{feynman} with the current $J_\mu$ turned off can be written as
\begin{equation}\label{feynmanin}
Z(0)=\int DG\int DA_\mu e^{i S_y}\Delta_{FP}(A^{bG}(x))  \delta[F_a(A^{bG}(x))-F_a(A^b(x))].
\end{equation}
The next task is to eliminate the Delta function, which has the gauge invariance property  $\Delta_{FP}(A^{bG}(x))=\Delta_{FP}(A^{b}(x))$. This procedure  is described in pages 228-237 in the book \cite{buchbinder2}, see also \cite{buchbinder}, \cite{ryder}.  By taking into account the representation \eqref{popin} a for the Faddev-Popov determinant, the resulting gauge fixed functional is
$$
Z(J)=\int DA^b_\mu  D\overline{\eta}^a D\eta^c \exp\{i S_y-\frac{i\lambda}{2}\int d^4x F_a(A^b_\mu)^2\}
$$
\begin{equation}\label{feynman2}
 \exp\{-i\int d^4y d^4x\,(\overline{\eta}^a (x)M_{ac}(x, y)\eta^c(y))+i\int d^4x\; J_\mu A^\mu\}.
\end{equation}
The action in the last expression corresponds to a lagrangian which contains the gauge fixing term
$$
L_{gf}=L_y-\frac{\lambda}{2}F_a(A^b_\mu)^2.
$$
The terms proportional to $M_{ab}(x,y)$ correspond to the Faddeev-Popov determinant. This expression will be important in the next sections.

The  Faddeev-Popov procedure can be generalized to non linear gauges \cite{weinberg}-\cite{bilal}, but for the present purposes the last formula is enough.
\subsection{Generalizations of gauge symmetry}

The expression \eqref{feynman2} for the Feynman path integral leads to the gauge fixed  action 
\begin{equation}\label{efectiva}
S_{gf}=\int \;d^4x\;  [L_y-\frac{\lambda}{2}F_a(A^b_\mu)^2-\overline{\eta}^aM_{ab} \eta^b].
\end{equation}
Here the operator $M_{ab}(x,y)$ usually has a delta such as in \eqref{tatu} has been converted in one depending on one variable $M_{ab}(x)$, in analogous fashion as in \eqref{identificatiion}. The last action, of course, is non gauge invariant, as it reflects an specific choice of gauge. An important task is to determine if there is a generalization of the gauge symmetry \eqref{variando},  involving $A^b_\mu$, $\eta^a$ and $\overline{\eta}^a$ leaving $S_{gf}$ invariant. Denote this potential new symmetry as $\delta_n$, where the subscript $n$ emphasizes  its novelty. Since the component $L_y$ involves only $A_\mu$ and is gauge invariant, it may be natural, although not necessarily true,  to expect that $\delta_n A^a_\mu=D_\mu^{ab}\theta^b$. By employing this ansatz, the goal is to determine how $\delta_n$ acts on the remaining fields $\eta^a$ and $\overline{\eta}^a$. This has to be determined with the requirement that the variation of the remaining terms in the gauge-fixed action,  
\begin{equation}\label{sobran}
L_g=-\frac{\lambda}{2}F_a(A_\mu^b)^2-\overline{\eta}^a M_{ab} \eta^b,
\end{equation}
vanishes. A generic variation, without specifying explicitly $\delta_n$ at this stage, is given by
\begin{equation}\label{sobran2}
\delta_n L_g=-\lambda F^a \delta_n F_a-\delta_n\overline{\eta}^a M_{ab} \eta^b-\overline{\eta}^a \delta_n(M_{ab} \eta^b).
\end{equation}
Due to the identification \eqref{identificatiion} it is clear that 
$$
M_{ab}\eta^b=\phi^\mu_{ac} D_{\mu}^{cb}\eta^b.
$$ On the other hand,  the formula \eqref{noabel} leads to
$$
\delta F^a(A_\mu^b)=M_{ab}\theta_b(x)=\phi^\mu_{ac} D_{\mu}^{cb}\theta^b.
$$
The last two formulas are remarkably similar. From this similarity, it is seen that the identification  $\theta^a =\eta^a\epsilon$
together with 
\begin{equation}\label{atento1}
\delta_n\overline{\eta}^a=-\lambda F^a \epsilon,
\end{equation}
ensures that the first two terms in \eqref{sobran2} cancel.  The identification $\theta^a =\eta^a\epsilon$ is basically describing a gauge transformation with the ghost $\eta^a$ playing the role of a gauge parameter.  The corresponding gauge transformation is 
\begin{equation}\label{atento2}
\delta_n A^a_\mu=D_\mu^{ab}\eta^b\epsilon.
\end{equation}
Here the proportionality factor $\epsilon$ is assumed to be a real parameter.  Note, however, that if $\epsilon$ is chosen as a Grassmann variable, which implies that $\epsilon^2=0$ and $\eta^a\epsilon=-\epsilon \eta^a$, then the transformation become
\begin{equation}\label{atento3}
\delta_n\overline{\eta}^a=\lambda F^a \epsilon,\qquad
\delta_n A^a_\mu=D_\mu^{ab}\eta^b\epsilon.
\end{equation}
The main difference between \eqref{atento1}-\eqref{atento2} with \eqref{atento3} is a change of a plus sign with a minus  one.

The transformations \eqref{atento1}-\eqref{atento2} or \eqref{atento3} given above still do not describe a symmetry, unless   $\delta(M_{ab} \eta^b)=0$. Since these transformations \eqref{atento1}-\eqref{atento2} and \eqref{atento3} affect the anti-ghost field $\overline{\eta}^a$ and the gauge field
$A_\mu^a$, the requirement  $\delta(M_{ab} \eta^b)=0$ has to be employed to fix the transformation of the ghost $\eta^a$, which is still missing.  An attempt could be  $\delta_n \eta^a=0$. The issue is that, in a generic situation, $M_{ab}$ contains the gauge field $A^a_\mu$ inside the covariant derivative $D_\mu^{cb}$, rendering this ansatz inefficient. This problem, of course, is avoided if $c_{abc}=0$, which corresponds to abelian models.  In this case $D_\mu^{ab}$ reduces to $\partial_\mu$, which is obviously $A^\mu$ independent. This implies that, for abelian models, a generalization of the gauge symmetry is described by
\begin{equation}\label{abel}
\delta_n A_\mu=\partial_\mu \eta,\qquad \delta_n\eta=0,\qquad\delta_n\overline{\eta}=-\lambda F.
\end{equation}
For the non abelian case this does not work. It is however possible to find an action $\delta_n \eta^a=0$ such that $\delta_n(M_{ab} \eta^b)$ vanishes. In a generic situation, the only constraint available for the gauge algebra is the Jacobi identity
$$
c_{abc}c_{cde}+c_{adc}c_{ceb}+c_{aec}c_{cbd}=0.
$$
The task is to determine if this constraint can be employed for the desired task. Assume first that $\epsilon$ is real and put $\epsilon=1$ by simplicity. Then 
\begin{equation}\label{nabel}
\delta_n A_\mu=D_\mu^{ab}\eta^b,\qquad \delta_n\overline{\eta}^a=-\lambda F^a,
\end{equation}
while $\delta_n\eta^a$ is yet to be specified by the requirement $\delta_n(M_{ab} \eta^b)=0$. By employing the formula \eqref{noabel}, one directly arrives to
$$
\delta_n(M_{ab} \eta^b)=\phi_{ac}^\mu \delta_n(D_\mu^{cb}\eta^b).
$$
This implies that the quantity that must vanish is
\begin{equation}\label{musvanish}
\delta_n(D_\mu^{ab}\eta^b)=\delta_n(\partial_\mu \eta^a-gc_{abc}\eta^b A^c_\mu )=\partial_\mu \delta_n\eta^a-gc_{abc}\eta^b(\delta_nA^c_\mu)-gc_{abc} \delta_n \eta^b A_\mu^c,
\end{equation}
by choosing an appropriate ansatz for $\delta_n\eta^a$.  By use of \eqref{nabel} the last expression becomes 
$$
\delta_n(D_\mu^{ab}\eta^b)=\partial_\mu \delta_n\eta^a-gc_{abc}\eta^b(\partial_\mu \eta^c-gc_{cde}\eta^d A^e_\mu)-gc_{abc}\delta_n \eta^bA^c_\mu .
$$
The vanishing of the terms with $\partial_\mu$ in the above formula leads to
\begin{equation}\label{ansatz}
\delta_n\eta^a=\frac{gc_{abc}}{2}\eta^b \eta^c,
\end{equation}
where the anti commutation property of $\eta^a\eta^b=-\eta^b\eta^a$ was employed in order to delete the derivative $\partial_\mu$ and consequently obtaining  this relation. However, the terms that do not contain $\partial_\mu$  has to cancel each other  as well. Fortunately they do, due to the Jacobi identity. To see this, observe that the remaining terms not proportional to $\partial_\mu$, denoted with the letter  $O$ of obstruction  in the following
$$
O=g^2c_{abc}c_{cde} A^e_\mu \eta^b \eta^d-gc_{abc}A^c_\mu \delta_n \eta^b,
$$
become with the help of \eqref{ansatz} 
$$
O=g^2c_{abc}c_{cde} A^e_\mu \eta^b \eta^d- \frac{g^2}{2}c_{abc}c_{bde}A^c_\mu\eta^d \eta^e.
$$
After re-denominating indices this may be written as
$$
O=g^2c_{abc}c_{cde} A^e_\mu \eta^b \eta^d+\frac{g^2}{2}c_{abc}c_{cde}A^b_\mu\eta^d \eta^e.
$$
In order to take the ghost and gauge fields as common factor, this may be expressed in the following form
$$
O=\frac{g^2}{2}(2c_{abc}c_{cde}+c_{aec}c_{cbd})A^e_\mu\eta^b \eta^d=\frac{g^2}{2}(c_{abc}c_{cde}-c_{adc}c_{cbe}+c_{aec}c_{cbd})A^e_\mu\eta^b \eta^d.
$$
Here the property $\eta^a\eta^b=-\eta^b\eta^a$ was employed. As $c_{cbe}=-c_{ceb}$ it is seen from the last expression that
$$
O=\frac{g^2}{2}(c_{abc}c_{cde}+c_{adc}c_{ceb}+c_{aec}c_{cbd})A^e_\mu\eta^b \eta^d=0,
$$
the last conclusion is due to the Jacobi identity for $c_{abc}$. Therefore the gauge fixed action \eqref{efectiva}
is invariant under
\begin{equation}\label{brstreal}
 \delta_\epsilon A_\mu=D_\mu^{ab}\eta^b\epsilon ,\qquad \delta_\epsilon\eta^a=\frac{gc_{abc}}{2}\eta^b \eta^c\epsilon,\qquad \delta_\epsilon\overline{\eta}^a=-\lambda F^a\epsilon,
\end{equation}for $\epsilon$ a real parameter.  The above calculation may be repeated for $\epsilon$ a Grassmann variable, which implies that $\epsilon\eta ^a=-\eta^a\epsilon$. This is the most classical calculation and the result is \cite{brst}
\begin{equation}\label{brstgrassmann}
 \delta_\epsilon A_\mu=D_\mu^{ab}\eta^b\epsilon ,\qquad \delta_\epsilon \eta^a=-\frac{gc_{abc}}{2}\eta^b \eta^c\epsilon,\qquad \delta_\epsilon\overline{\eta}^a=\lambda F^a\epsilon.
\end{equation}
Both versions \eqref{brstreal} and \eqref{brstgrassmann} are symmetries of the action \eqref{efectiva}.  In both cases 
$\delta_\epsilon(D_\mu^{ab}\eta^b)=0$, as this was the condition required to determine $\delta_\epsilon \eta^a$ in the procedure above. Also in both cases $\delta_\epsilon(c_{abc}\eta^b \eta^c)=0$.  This can be checked as follows. For $\epsilon$ a commuting variable there is no loss of generality in choosing  $\epsilon=1$ in order to find that
$$
\delta_\epsilon(c_{abc}\eta^b \eta^c)=c_{abc}\delta_\epsilon(\eta^b) \eta^c+
c_{abc}\eta^b\delta_\epsilon(\eta^c)=\frac{g}{2}c_{abc}c_{bde}\eta^d \eta^e\eta^c+\frac{g}{2}c_{abc}c_{cde}\eta^b\eta^d \eta^e
$$
$$
=\frac{g}{2}c_{abc}c_{bde}\eta^d \eta^e\eta^c+\frac{g}{2}c_{acb}c_{bde}\eta^c\eta^d \eta^e=\frac{g}{2}c_{abc}c_{bde}\eta^d \eta^e\eta^c+\frac{g}{2}c_{acb}c_{bde}\eta^d \eta^e\eta^c=0.
$$
In the last step the fact that $c_{acb}=-c_{abc}$ was employed.  For $\epsilon$ a Grassman variable instead the fact that $\epsilon\eta ^a=-\eta^a\epsilon$ leads to
$$
\delta_\epsilon(c_{abc}\eta^b \eta^c)=-\frac{g}{2}c_{abc}c_{bde}\eta^d \eta^e\epsilon \eta^c-\frac{g}{2}c_{abc}c_{cde}\eta^b\eta^d \eta^e\epsilon=\frac{g}{2}c_{abc}c_{bde}\eta^d \eta^e\eta^c\epsilon-\frac{g}{2}c_{abc}c_{cde}\eta^d \eta^e\eta^b\epsilon
$$
$$
=\frac{g}{2}c_{abc}c_{bde}\eta^d \eta^e\eta^c\epsilon-\frac{g}{2}c_{acb}c_{bde}\eta^d \eta^e\eta^c\epsilon=gc_{abc}c_{bde}\eta^d \eta^e\eta^c\epsilon=0.
$$
The last step follows directly from the Jacobi identity.

In the following, the version \eqref{brstgrassmann} will be employed, which corresponds to a Grassmann parameter $\epsilon$. This is the celebrated BRST symmetry for gauge theories \cite{brst}. It is customary to denote the BRST variation of the field $F$  with the symbol $s F$, with the parameter $\epsilon$ omitted. This definition, when applied to the fundamental fields of Yang-Mills theories, is given by
\begin{equation}\label{s}
sA_a^\mu=D_{ab}^\mu \eta^b,\qquad s \eta^a=-\frac{g}{2}c_{abc}\eta^b \eta^c,\qquad s \overline{\eta}^a=\lambda F_a(A).
\end{equation}
The last discussion has shown that $s^2 A_b^\mu=s^2\eta^a=0$. Therefore the action of $s$ is nilpotent on those fields.  This is not the case however for the action on $\overline{\eta}^a$. Nevertheless, the partial nilpotency of $s$ plays a crucial role in establishing certain gauge identities for the generating functionals of the theory.  

A final important comment is that given a function $F(A^a_\mu, \eta^b)$, then
$s^2 F=0$ due to the above derived nilpotency conditions. In fact
$$
\delta F=\frac{\partial F}{\partial A_\mu^a}s A_\mu^a \epsilon +\frac{\partial F}{\partial \eta^b}s \eta^b \epsilon,
$$
which implies that
$$
\delta s F=\frac{\partial F}{\partial A_\mu^a \partial A_\nu^b} \delta A_\nu^bs A_\mu^a  +\frac{\partial F}{\partial \eta^a \partial \eta^b}s \eta^b\epsilon  s \eta^ a +\frac{\partial F}{\partial A_\mu^a \partial \eta^b}(s \eta^b\epsilon s A_\mu^a  +s A_\mu^a\epsilon s \eta^ b).
$$
The first term vanish since $\delta A_\mu^a\sim D^{ab}_\mu\eta^b$ and this term is a symmetric times anti symmetric expression due to the anti commutation property of $\eta^a$.  The last two terms vanish since in the first term the parameter $\epsilon$ goes through $sA_\mu^a$, which is linear expression in $\eta^a$ due to the BRST definition \eqref{brstgrassmann}, and through the quadratic expression $s\eta^a$ in $\eta^\alpha$ in the second term. Therefore, one term picks a minus sign and the other not. Furthermore $s A_\mu^a s\eta^b=s\eta^b sA_\mu^a$ as one term is linear and the other quadratic in the ghosts $\eta^a$ due to the definition \eqref{brstgrassmann}, and this shows the cancellation of the last two terms.  The only term that remains is the second, that is 
\begin{equation}\label{ufaa}
s^2 F=\frac{\partial F}{\partial \eta^a \partial \eta^b}s \eta^b  s \eta^ a.
\end{equation}
As \eqref{brstgrassmann} shows that $s\eta^a$ is quadratic in the ghost, it is clear that $s \eta^b  s \eta^ a=s \eta^a  s \eta^b$. However, the subtle point is that  the partial derivative with respect to the ghost is not symmetric. This can be seen as follows. Thinking the derivative of a ghost as an ordinary derivative will lead to the conclusion
$$
\frac{\partial}{\eta^a}(\eta^a\eta^b)=\eta^b=-\frac{\partial}{\eta^a}(\eta^b\eta^a)=-\eta^b, 
$$
which is absurd. This false statement is avoided by assuming that the derivative has a definite sign, chosen as positive, when acting in the first Grassmann variable $\eta^a$ and chosen negative  when acting on the second variable $\eta^b$. This derivative is then anti-symmetric, that is 
$$
\frac{\partial F}{\partial \eta^a \partial \eta^b}=-\frac{\partial F}{\partial \eta^b\partial \eta^a}.
$$
In these terms it is clear that \eqref{ufaa} vanishes and therefore
\begin{equation}\label{ufaa2}
s^2 F=0,
\end{equation}
for every $F(A_\mu^a, \eta^b)$. Thus, nilpotency is preserved for all the functions of the gauge field $A_\mu^a$ and the ghost $\eta^a$.

\subsection{Connected Green functions and generalizations of BRST symmetry}
As it is well known, the gauge fixed Feynman integral \eqref{feynman2} is a fundamental tool in order to calculate to generic correlation functions. On the other hand, it   is known that the action   $W(J, \xi, \overline{\xi})$  defined by
$$
e^{i W(J, \xi, \overline{\xi})}=\int DA_\mu^a D\eta^a D\overline{\eta}^a \exp\{i\int \;d^4x\;  [L_y-\frac{\lambda}{2}F_a^2-\overline{\eta}^aM_{ab}\eta^b+J^\mu_a\cdot A^a_\mu+\overline{\xi}_a\eta^a+\xi_a\overline{\eta}^a]\},
$$
is the generator for connected Green functions. Here, $J$ is a source current coupled to the gauge field, while $\xi$ and $\overline{\xi}$ are new sources that couple to the Faddeev-Popov anti-ghost and ghost respectively. The definition given above is equivalent to the following one
$$
i W(J, \xi, \overline{\xi})=\log Z(J, \xi, \overline{\xi}).
$$
For gauge theories, correlation functions such as $<A_\mu^a(x)A_\nu^b(y)>$ are  not gauge invariant, and for this reason it is customary to employ Wilson loops.  However, for studying identities related to the BRST invariance of the model, it is sometimes more convenient to introduce a generalization of the action $W(J, \xi, \overline{\xi})$, which will be denoted here as  $G(J, \xi, \overline{\xi}, K, L)$. The main difference is the introduction of two currents $K_\mu^a$ and $L_a$ which couple to the nilpotent part of the BRST variations  \eqref{s}. In other words, the functional $G(J, \xi, \overline{\xi}, K, L)$ is defined through the formula \cite{kluberg}
$$
e^{i G(J, \xi, \overline{\xi}, K, L)}=\int DA_\mu^a D\eta^a D\overline{\eta}^a \exp\{i\int \;d^4x\;  [L_y-\frac{\lambda}{2}F_a^2-\overline{\eta}^aM_{ab}\eta^b+K_\mu^asA^\mu_a-L_a s\eta^a
$$
\begin{equation}\label{conexa}
+J^\mu_a\cdot A^a_\mu+\overline{\xi}_a\eta^a+\xi_a\overline{\eta}^a]\},
\end{equation}
where $s A$ and $s\eta$ are the BRST variations of these fields introduced in \eqref{s}.  This is the definition given in page 597 of \cite{itzykson}. It is not difficult to see that the functional $G(0,0,0, K, L)$  is invariant under a BRST transformation  due to the nilpotency condition $s^2 A_b^\mu=s^2\eta^a=0$ described in the previous subsection.  If furthermore the new currents are turned off  then the following identification takes place$$W(J, \xi, \overline{\xi})=G(J, \xi, \overline{\xi}, 0, 0 ).$$
The last formula implies that  the knowledge of $G(J, \xi, \overline{\xi}, K, L)$ leads the complete determination of $W(J, \xi, \overline{\xi})$.

At this point it is interesting to determine the dimensions of the fields of the action inside the path integral \eqref{conexa}. This action reads explicitly  
$$
S_{gf}( A, \eta, \overline{\eta},J, \xi, \overline{\xi}, K, L)=\int \;d^4x\;  [L_y-\frac{\lambda}{2}F_a^2-\overline{\eta}^a\phi^{ab}_\mu D^\mu_{ab}\eta^b+K_\mu^aD_{ab}^\mu \eta^b+\frac{g}{2}L_a c_{abc}\eta^b \eta^c
$$
\begin{equation}\label{apagada}
+J^\mu_a A^a_\mu+\overline{\xi}_a\eta^a+\xi_a\overline{\eta}^a].
\end{equation}
The mass dimensions of the fields are directly seen from this action. They are
\begin{equation}\label{masa}
[A_\mu^a]=[\eta^a]=[\overline{\eta}^a]=1, \qquad [J_a^\mu]=[\xi_a]=[\overline{\xi}_a]=3,\qquad [K_\mu^a]=[L_a]=2.
\end{equation}
The dimensions of the Faddeev-Popov ghosts are not unique, the only constraint for them is that $[\eta^a]+[\overline{\eta}^a]=2$. However, the choice above is valid and is the one to be employed below. 

Another important issue is how the quantities above transform under Lorentz transformation. The fields $A_\mu^a$ is a vector, while the ghost $\eta^a$ and $\overline{\eta}^a$ are scalars. The lagrangian should be a Lorenz scalar, which means that $J_a^\mu$ is a vector, $\xi^a$ and $\overline{\xi}^a$ are scalars, $K_\mu^a$ is a vector and $L_a$ is a scalar. 

In addition, the action is invariant under the global symmetry, usually known as ghost symmetry, given by
\begin{equation}\label{ghostsy}
\eta^a\to e^{i\alpha}\eta^a,\qquad  \overline{\eta}^a\to e^{-i\alpha}\overline{\eta}^a, \qquad \xi_a\to e^{-i\alpha}\xi_a,\qquad  \overline{\xi}_a\to e^{i\alpha} \overline{\xi}_a,
\end{equation}
$$
K_\mu^a\to e^{-i\alpha}K_\mu^a,\qquad L_a\to e^{-2i\alpha} L_a,
$$
where the phase $\alpha$  is independent on the space time coordinates. The remaining fields $A_\mu^a$ and $J_a^\mu$ are unchanged under this action. The charge of these fields is known as ghost number, denoted by $Q$ and can be read from the last formula
\begin{equation}\label{Q}
Q(A_\mu^a)=Q(J_a^\mu)=0,\qquad Q(L_a)=-2,
\end{equation}
$$
Q(\eta^a)=-Q(\overline{\eta}^a)=-Q(\xi_a)=Q(\overline{\xi}_a)=- Q(K_\mu^a)=1.
$$
The action $S( A, \eta, \overline{\eta},J, \xi, \overline{\xi}, K, L)$  has zero ghost number, since it is invariant under this global symmetry. In addition, the BRST action \eqref{s} increases the ghost number by one. In this discussion it is assumed that the action of $s$ does not modify the introduced currents $K_\mu$ and $L_a$.
\\

\emph{Three important generalizations}
\\

It is instructive at this point to address three important issues. First, to answer which is the most general action $S_{gene}( A, \eta, \overline{\eta}, K, L)$ which satisfies the requirements of being invariant under $s$, having zero ghost number,  being  Lorentz invariant and with mass dimensions less or equal than four. The last requirement is due to renormalizability issues. This question is referred to the situation in which  $J_\mu^a=\xi^a=\overline{\xi}^a=0$, as the BRST invariance applies to the part of the lagrangian with these currents turned off. 
\\

\emph{First problem:} From \eqref{masa}, it is seen that the currents $K_\mu^a$ and $L_a$ have mass dimension two, so they may appear at most in a quadratic term.  Thus, they may appear at most in terms such as
$$
L_{KK}\sim K^a_\mu K^\mu_a.
$$
However, this term has ghost number $-2$ and therefore is forbidden. The same argument goes with the current $L_a$ or mixtures between both ghost currents. Therefore the lagrangian is at most linear in $K_\mu^a$ and  $L_a$. A term of the form 
$$
L_L=L_a \partial^\mu A_\mu^a,
$$
is Lorenz invariant and has the correct mass dimensions, however, it does not have ghost number zero. 
A term of the following type
$$
L=c K_a^\mu A_\mu^a,
$$
have mass dimension less than four and are Lorenz invariant, but it does not have ghost number zero either. After some reasoning and by writing out the possible terms, it is concluded  that the action \eqref{apagada} is quite general. Specifically, when current $J_\mu^a$, $\xi^a$ and $\overline{\xi}^a$ are turned off, it is the most general BRST invariant action that fulfills all the above requirements, up to a field redefinition. Explicitly
\begin{equation}\label{apagada33}
S_{gene}( A, \eta, \overline{\eta}, K, L)=\int \;d^4x\;  [L_y-\frac{\lambda}{2}F_a^2-\overline{\eta}^a\phi^{ab}_\mu D^\mu_{ab}\eta^b+K_\mu^aD_{ab}^\mu \eta^b+\frac{g}{2}L_a c_{abc}\eta^b \eta^c].
\end{equation}
\emph{Second problem:} The second important question to answer is, which is the most general symmetry $s_n$ generalizing \eqref{s} increasing the ghost number by one, and which  is nilpotent when acting on  $A_\mu^a$ and $\eta^a$. The condition of increasing the ghost number by one is because it is evident that the standard BRST symmetry  \eqref{s} has precisely this property. Dimensional analysis and ghost increase number requirement indicate that  
\begin{equation}\label{s2}
s_nA^a_\mu=a_{ab}\partial_\mu \eta^b-gd_{abc}A^b_\mu\eta^c ,\qquad s_n \eta^a=-\frac{g}{2}e_{abc}\eta^b \eta^c.
\end{equation}
The nilpotency condition on $\eta^a$ leads to
$$
 \delta_\epsilon(s_n\eta^a)=-\frac{g}{2}e_{abc}e_{cde}\eta^b \eta^d\eta^e\epsilon.
$$
The vanishing of this contribution leads to $e_{abc}e_{cde}+$ (cyclic permutations)=0. This implies that $e_{abc}$
satisfy a Jacobi like identity, therefore it can be interpreted as the structure constant of a suitable Lie algebra. In addition, the remaining nilpotency condition is
$$
\delta_\epsilon(s_nA^a_\mu)=a_{ab}\partial_\mu \delta_\epsilon\eta^b-gd_{abc}\delta_\epsilon(A^b_\mu)\eta^c-gd_{abc}A^b_\mu\delta_\epsilon(\eta^c)
$$
$$
=-\frac{g}{2}e_{bfc}e_{cde}a_{ab}\partial_\mu (\eta^f \eta^d\eta^e)\epsilon+gd_{abc}(a_{bd}\partial_\mu \eta^d-gd_{bde}A^d_\mu\eta^e)\eta^c\epsilon
+\frac{g^2}{2}d_{abc}e_{cde}A_\mu^b\eta^d \eta^e\epsilon.
$$
The first term vanish due to the Jacobi identity for $e_{abc}$. The vanishing of the term proportional to $\partial_\mu$ indicate that
$$
d_{abc} a_{bd}=0,
$$
while the vanishing of the terms proportional to $g^2$ results in the following relation
$$
d_{ace}d_{cbd}-d_{acd}d_{cbe}=e_{cde}d_{abc}.
$$
A solution of the last two equations can be found as follows. By defining the matrix $(T^a)_{bc}=id_{bca}$ the last equation becomes
$$
[T^a, T^b]=ie_{cab}T^{c}.
$$
By interpreting  $e_{bec}$  as structure constants of a Lie algebra, it follows that $T^a$ lies in its adjoint representation. In other words $d_{abc}=e_{abc}$. Finally, the fact that $d_{abc} a_{bd}=0$ implies that $a_{ab}$ commutes with all the generators of the algebra. If the algebra is simple, that means that $a_{ab}=a \delta_{ab}$. This will be the case to be considered below. Therefore, the generalization of BRST symmetry for simple Lie algebras reads as follows
\begin{equation}\label{s3}
s_nA^a_\mu=a\partial_\mu \eta^a-gd_{abc}A^b_\mu\eta^c ,\qquad s_n \eta^a=-\frac{g}{2}d_{abc}\eta^b \eta^c,\qquad s_n\overline{\eta}^a=\lambda F^a(A_\mu^b).
\end{equation}
By redefining the fields $\eta^a\to a \eta^a$, $d_{abc}\to   a^{-1}d_{abc}$ and by leaving $A_\mu^a$ and $\overline{\eta}^a$ unchanged, the last  symmetry \eqref{s3} takes the same form as the BRST one given in  \eqref{s}.  

The fact that the BRST symmetry has been rediscovered if the fields are redefined implies that the most general action invariant under \eqref{s3} with ghost number zero and Lorenz invariant, with dimension four or less, is given by
\begin{equation}\label{generaltotal}
S_{gen}( A, \eta, \overline{\eta}, K, L)=\int  d^4 x[l(A)-\frac{\lambda}{2}F_a^2+(K^{\mu}_{a}-(\overline{\eta}\phi^\mu)_{a})\Delta_{\mu a b}\eta_{b}+\frac{g}{2}d_{abc}L_{a} \eta_{b} \eta_{c}],
\end{equation}
where the following generalization of the covariant derivative has been introduced
$$
\Delta_{\mu a b}=a \partial_\mu \delta_{ab}-g d_{abc} A^b_{ \mu},
$$
and a redefinition of the current $\overline{\xi}^a\to a\overline{\xi}^a$ has been made. The term involving $(\overline{\eta}\phi^\mu)_{a}$ may be multiplied by a constant, but it can be adsorbed in the definition of $\overline{\eta}^a$. The function $l(A)$ is the most general gauge invariant function of mass four of less, constructed from the gauge field $A_\mu^a$ and its derivatives.
\\

\emph{The third problem:} The third and last problem is to classify  the most general action   $S_{nbrst}(A, \eta, \overline{\eta}, K, L)$   which has zero ghost number, it is invariant under Lorenz transformations, has dimension four or less due to renormalizability, and depends on the fields $K^{\mu}_{a}$ and $\eta_a$ only through the linear combination $K^{\mu}_{a}-(\overline{\eta}\phi^\mu)_{a}$. The main difference with the above analysis is that BRST invariance is not required. Although this last question may seem less motivated than the above ones, this analysis will be useful when dealing with Slavnov-Taylor identities.
As the combination $K^{\mu}_{a}-(\overline{\eta}\phi^\mu)_{a}$ has ghost number $-1$ due to \eqref{Q} and mass dimension $2$ due to \eqref{masa}
the only way that it may appear is
\begin{equation}\label{casigeneraltotal}
S_{nbrst}(A, \eta, \overline{\eta}, K, L)=\int  d^4 x[l(A^b_\mu)+(K^{\mu}_{a}-(\overline{\eta}\phi^\mu)_{a})\Delta_{\mu}^{ a b}\eta_{b}+\frac{1}{2}d_{abc}L_{a} \eta_{b} \eta_{c}],
\end{equation}
where now
$$
\Delta_{\mu}^{ a b}=\alpha Q^ {ab} \partial_\mu-\beta g c_{abc} A^c_{ \mu}.
$$
This is the most general coupling that has zero ghost number and is Lorentz invariant.  If the matrix $Q^{ab}$ is invertible, then the redefinition $$\eta^{\prime a}=Q^{ab}\eta^b,\qquad c^{\prime}_{adc}=c_{abc}(Q^{-1})^{bd},\qquad d^{\prime}_{abc}=d_{ade}(Q^{-1})^{db}(Q^{-1})^{ec},$$
leads to 
\begin{equation}\label{generaltotaldes}
S_{nbrst}( A, \eta, \overline{\eta}, J, \xi, \overline{\xi}, K, L)=\int  d^4 x[l(A_\mu^a)+(K^{\mu}_{a}-(\overline{\eta}\phi^\mu)_{a})\Delta_{\mu a b}\eta_{b}+\frac{g}{2}d_{abc}L_{a} \eta_{b} \eta_{c}],
\end{equation}
where the primes where deleted by simplicity and 
$$
\Delta_{\mu a b}=\alpha \partial_\mu \delta_{ab}-\beta g c_{abc} A^c_{ \mu},
$$
Here the hypothesis that the action depends on $K^{\mu}_{a}-(\overline{\eta}\phi^\mu)_{a}$ was employed. The formula \eqref{generaltotaldes} strongly resembles  to \eqref{generaltotal}. However, the difference  $a$ and $\beta$ are not necessarily equal to one. In addition the constants $d_{abc}$
are not related a priori to the gauge ones $c_{abc}$. As before $l(A_\mu^a)$ has dimensions four or less and contains derivatives of the field, but it can not be assumed that is gauge invariant. The utility of this expression will be clear soon.  Note that this formula is very briefly discussed in page 602 of reference \cite{itzykson}.

In order to close this section, it should be mentioned that the references \cite{weinberg}-\cite{bilal}
study similar questions, with the difference that they use an enlarged version of the BRST symmetry with new auxiliary fields, nilpotent for all of them. These references employ the Nakanishi-Lauptrup formalism \cite{nakanishi}-\cite{lauptrup} . The action is expressed as the sum of the Yang-Mills one and a pure BRST term $s F$ with $F$ some function of the fields. The BRST invariance is easier to be proved with this enlarged field content.  Furthermore, those expositions are adapted to more general gauges, and this is of clear interest.  Another formalism of particular interest is the Batalin-Vilkovisky one \cite{batalin1}-\cite{batalin2}, which is valid even for open symmetry algebras. The present approach is minimalist, and for this reason this general and important formulation is skipped.  

\subsection{The effective action and BRST symmetry}
It is well known that the  functional  $G(J, \xi, \overline{\xi}, K, L)$ defined in \eqref{conexa} is a fundamental tool for the calculation of the mean values
\begin{equation}\label{mean}
i<A^a_\mu>=\frac{\delta G}{\delta J_a^\mu},\qquad i<\eta^a>=\frac{\delta G}{\delta \overline{\xi}_a},\qquad i<\overline{\eta}^a>=-\frac{\delta G}{\delta \xi_a}.
\end{equation}
These mean values at the moment correspond to the sources $J_a^\mu, \xi^a, \overline{\xi}^a, K_a^\mu, L_a$ turned on. Usually, in physical applications, the mean values of interest correspond to  all the derivatives of  $G(J, \xi, \overline{\xi}, K, L)$ evaluated at $J_a^\mu= \xi^a= \overline{\xi}^a= K_a^\mu= L_a=0$, denote this quantities as $<A_\mu^a>_c$, $<\eta^a>_c$ and $<\overline{\eta}^a>_c$. The subscript $c$ indicates that these quantities are, in some sense, close to a classical field. Explicitly, they are given by
$$
i<A^a_\mu>_c=\frac{\delta G}{\delta J_a^\mu}\bigg|_{J_a^\mu= \xi^a= \overline{\xi}^a= K_a^\mu= L_a=0},\qquad i<\eta^a>_c=\frac{\delta G}{\delta \overline{\xi}_a}\bigg|_{J_a^\mu= \xi^a= \overline{\xi}^a= K_a^\mu= L_a=0},
$$
\begin{equation}\label{meandos}
i<\overline{\eta}^a>_c=-\frac{\delta G}{\delta \xi_a}\bigg|_{J_a^\mu= \xi^a= \overline{\xi}^a= K_a^\mu= L_a=0}.
\end{equation}
The point is that the fields $A_\mu^a$, $\eta^a$ and $\overline{\eta}^a$ are not measurable quantities, due to the uncertainty principle.  Instead, their mean values are. The exact mean values, of course, are difficult to be calculated, as it requires a complete knowledge of $G(J, \xi, \overline{\xi}, K, L)$. An interesting question is if there exists a functional $\Gamma( <A>, <\eta>, <\overline{\eta}>, K, L)$ such that its classical equations of motion leads to those mean values \cite{goldstone}.  In other words, the solutions of the equations of motion 
\begin{equation}\label{legendre}
\frac{\delta \Gamma}{\delta <A^a_\mu>}=-J_a^\mu,\qquad -\frac{\delta \Gamma}{\delta <\overline{\eta}^a>}=-\xi_a,\qquad \frac{\delta \Gamma}{\delta <\eta^a>}=\overline{\xi}_a,
\end{equation}
are expected to correspond precisely to the mean values $<A_\mu^a>$, $<\eta^a>$ and $<\overline{\eta}^a>$ as defined in \eqref{mean}. Note that these values have to be interpreted as those corresponding to fixed values of the currents $J_a^\mu, \xi^a, \overline{\xi}^a, K_a^\mu, L_a$.  In particular, the equations of motion at zero current 
\begin{equation}\label{legendredos}
\frac{\delta \Gamma}{\delta <A^a_\mu>}=0,\qquad -\frac{\delta \Gamma}{\delta <\overline{\eta}^a>}=0,\qquad \frac{\delta \Gamma}{\delta <\eta^a>}=0,
\end{equation}
should have $<A_\mu^a>_c$, $<\eta^a>_c$ and $<\overline{\eta}^a>_c$ as solution.

The formulas \eqref{mean} and \eqref{legendre} show that the functionals $G(J, \xi, \overline{\xi}, K, L)$ and   $\Gamma( A, \eta, \overline{\eta}, K, L)$ are related by a Legendre transform 
$$
\Gamma( <A>, <\eta>,<\overline{\eta}>, K, L)=-i G(J, \xi, \overline{\xi}, K, L)
$$
\begin{equation}\label{quantumaction}
-\int d^4 x[J_a^\mu  <A_\mu^a>+\overline{\xi}_a<\eta^a>+\xi_a<\overline{\eta}^a>].
\end{equation}
The functional $\Gamma( <A>, <\eta>,<\overline{\eta}>, K, L)$ is known as the quantum effective action \cite{zinnjustin}-\cite{zinnjustin3}.  Of course, the explicit determination of this action is a challenging task, as the mean values and $G(J, \xi, \overline{\xi}, K, L)$ are typically difficult to be  computed. 

There are several important properties of the quantum effective action discussed in the literature, for instance in pages 154-159 of \cite{nastase}.  One is that $\Gamma(<A>, <\eta>, <\overline{\eta}>, K, L)$ generates the amputated $n$-point functions $\Gamma^n_{i_1...i_n}
  (x^1,...,x^n)$, that is, the Green functions with the external lines amputated, through the formula
\begin{equation}\label{amputado}
  \Gamma(<A>, <\eta>, <\overline{\eta}>, K, L)=\sum_{n=0}^\infty \frac{1}{n!}\int d^4x^1...d^4x^n\Gamma^n_{i_1...i_n}
  (x^1,...,x^n)\phi_{i_1}(x_1)...\phi_{i_m}(x^n).
\end{equation}
Here $\phi_i$ denotes any of the fields $<A>, <\eta>, <\overline{\eta}>, K, L$. 

There is another extremely useful  relation between $G$ and   $\Gamma$. The relation is 
\begin{equation}\label{conexa2}
e^{i G(J, \xi, \overline{\xi}, K, L)}=\int_{\text{C.T.D}} DA_\mu^a D\eta^a D\overline{\eta}^a \exp\{i\int \;d^4x\;  [\Gamma( A, \eta, \overline{\eta}, K, L)+J^\mu_a\cdot A^a_\mu+\overline{\xi}_a\eta^a
+\xi_a\overline{\eta}^a]\},
\end{equation}
where the notation C.T.D. stands for connected three diagrams. The physical meaning of the last formula is the following.  If functional form of the effective action $\Gamma( A, \eta, \overline{\eta}, K, L)$  has been obtained exactly, the functional $G(J, \xi, \overline{\xi}, K, L)$ can be found by using the effective action in place of the original action and by calculating Feynman diagrams only at tree level, without loops. 

This fact can be seen briefly as follows, details are for instance in pages 66 
-69 of \cite{weinberg} or in pages 135-137 of \cite{coleman}. Consider the quantity $G_\Gamma(J, \xi, \overline{\xi}, K, L)$ defined through
\begin{equation}\label{conexa22}
e^{i G_\Gamma(J, \xi, \overline{\xi}, K, L)}=\int DA_\mu^a D\eta^a D\overline{\eta}^a \exp\{\frac{i}{\hbar}\int \;d^4x\;  [\Gamma( A, \eta, \overline{\eta}, K, L)+J^\mu_a\cdot A^a_\mu+\overline{\xi}_a\eta^a
+\xi_a\overline{\eta}^a]\}.
\end{equation}
The difference between \eqref{conexa2} and \eqref{conexa22} is that the first only relates to the connected three diagrams, while the second concerns all the possible diagrams.  There is no need to write $<>$ in both formulas, as the path integral accounts for all the possible configurations. In all the formulas above except the last one, units employed are such that $\hbar=1$. In the last formula, $\hbar$ has been restored for some specific reason, which will be clarified shortly. The functional $ G_\Gamma(J, \xi, \overline{\xi}, K, L)$  in \eqref{conexa2} is the analog to $ G(J, \xi, \overline{\xi}, K, L)$ but with the gauged fixed action \eqref{apagada}
with sources $J=\xi=\overline{\xi}
=0$ replaced by $\Gamma( A, \eta, \overline{\eta}, K, L)$. Explicitly the replacement is
$$
S_{gf}(A, \eta, \overline{\eta}, K, L)=\int \;d^4x\;  [L_y-\frac{\lambda}{2}F_a^2+(K_\mu^a-\overline{\eta}^b\phi^{ba}_\mu )D^\mu_{ac}\eta^c+\frac{g}{2}L_a C_{abc}\eta^b \eta^c]
$$
\begin{equation}\label{source}
\longrightarrow \Gamma( A, \eta, \overline{\eta}, K, L).
\end{equation}
If it was possible to calculate exactly the effective action $\Gamma( A, \eta, \overline{\eta}, K, L)$, which is a task  hard to achieve, then the functional $G_\Gamma(J, \xi, \overline{\xi}, K, L)$ could be found perturbatively with the standard methods of Quantum Field Theory, by use of \eqref{conexa2}.  The introduction of $\hbar$ explicitly implies the multiplication of every propagator by $\hbar^{-1}$ and every vertex by $\hbar$. Then, every graph is multiplied by $\hbar^{I-V}$ with $I$ the number of internal lines and $V$ the number of vertices. The quantity $I-V$ is related to the number of loops $L$ in a graph by the relation $I-V=L-1$. Therefore, an expansion of the form
$$
G_\Gamma(J, \xi, \overline{\xi}, K, L)=\sum_{L=0}\hbar^{L-1} G^{(L)}_\Gamma(J, \xi, \overline{\xi}, K, L),
$$
is expected. The last formula implies that the term with $L=0$ is multiplied by a factor of $\hbar^{-1}$. It is possible to directly verify the classical limit by taking $\hbar G_\Gamma(J, \xi, \overline{\xi}, K, L)$ as $\hbar \to 0$ and applying a saddle point approximation, thereby obtaining the Ehrenfest statement in the small-$\hbar$ regime. This statement reduces to the desired result expressed in \eqref{conexa2}. More details can be consulted in \cite{weinberg}, \cite{coleman}.

It should be emphasized that the expansion in $\hbar$ given above is plausible, but not necessarily true.  There are examples in the literature where the $\hbar$ expansion fails \cite{holstein}. The validity of the last expansion will be taken as a working assumption,  and is true in several cases, but it can not be taken as a rigorous result.

Note that the definition \eqref{conexa} for  $G(J, \xi, \overline{\xi}, K, L)$ implies, in the classical limit $\hbar\to 0$, that $G(J, \xi, \overline{\xi}, K, L)$ is the Legendre transformation of the action $S_{gfs}(A, \eta, \overline{\eta}, K, L)$ defined in \eqref{source}.
On the other hand $\Gamma(A, \eta, \overline{\eta}, K, L)$ is the Legendre transform of $G(J, \xi, \overline{\xi}, K, L)$  for every $\hbar$ value, by the very definition \eqref{quantumaction}. This suggest that
\begin{equation}\label{expansion}
\Gamma(A, \eta, \overline{\eta}, K, L)=\sum_{n=0}  \hbar^n \Gamma_n(A, \eta, \overline{\eta}, K, L).
\end{equation}
In particular, the discussion above implies that the zero-th order is 
\begin{equation} \label{clasico}
 \Gamma_0(A, \eta, \overline{\eta}, K, L)=  S_{gf}(A, \eta, \overline{\eta}, K, L),
\end{equation}
where $S_{gf}(A, \eta, \overline{\eta}, K, L)$ is given in \eqref{source}. The last identity states that the zero order effective action coincides with the classical action,  as this corresponds to the classic limit.

\subsection{The Slavnov-Taylor equations and their implications}

In a generic situation, the amputated Green  functions $\Gamma^n_{i_1...i_n}
  (x^1,...,x^n)$ defined in \eqref{amputado} or even the functionals $ \Gamma_n(A, \eta, \overline{\eta}, K, L)$ defined in \eqref{expansion} are  expected to contain divergent parts which, in the present context, are cured by use of counter-terms in the action \eqref{conexa}. The number of counter terms required to cancel these divergences is likely to be finite, as the theory is renormalizable. However, the problem arises as to whether these cancellations preserve BRST invariance. For instance, for a non abelian theory, a possible counter term has the form
  $$
L_{\text{counter}}=Z (\partial_\mu A_\nu-\partial_\nu A_\mu)[A^\mu, A^\nu],
  $$
with $Z$ a renormalization constant. This term alone does break gauge invariance and with gauge fixing terms it may break BRST invariance as well.  

At first sight, there may be no reason to care about BRST symmetry or gauge symmetry. After all, gauge invariance is not expected to hold if the gauge has been fixed and BRST may seem an accidental symmetry that plays no role. Instead, BRST symmetry is an important invariance. The action of the BRST on the ghosts on the other hand eliminate longitudinal models and enforce the Gauss law constraints \cite{kaku}- \cite{peskin} for the gauge theory. The same feature holds for gravity theories, as described for instance in section 2.2 of  \cite{verlinde}.  For these reasons BRST symmetry is something that one may desire to preserve order by order.  Therefore, it is essential to understand the types of constraints that BRST invariance imposes on $\Gamma(A, \eta, \overline{\eta}, K, L)$. This is the subject of the so-called Slavnov-Taylor identities \cite{slavnov}-\cite{taylor}.

In general, the Slavnov-Taylor identities do arise due to the BRST invariance of the gauge fixed action $S_{gf}$ defined in \eqref{efectiva}. However, they do not restrict the functional $ \Gamma(A, \eta, \overline{\eta}, K, L)$ to be BRST invariant, which on the other hand is the desired feature. That is the point which is of interest, which is to  analyzed in the following.

The Slavnov-Taylor identities follow by inspecting the definition of the generating functional of connected Green functions $G(J, \xi, \overline{\xi}, K, L)$ defined in  \eqref{conexa}.  A closer look at this definition, which for linear gauges reads as follows,
$$
e^{i G(J, \xi, \overline{\xi}, K, L)}=\int DA_\mu^a D\eta^a D\overline{\eta}^a \exp\{i\int \;d^4x\;  [L_y-\frac{\lambda}{2}F_a^2+(K_\mu^a-\overline{\eta}^b\phi^{ba}_\mu )D^\mu_{ac}\eta^c-L_a s\eta^a
$$
\begin{equation}\label{conexass}
+J^\mu_a\cdot A^a_\mu+\overline{\xi}_a\eta^a+\xi_a\overline{\eta}^a]\},
\end{equation}
leads to the conclusion that the non-BRST invariant terms in the effective action are given by  
the source part of the Lagrangian
$$
L_{nbrst}=J^\mu_a\cdot A^a_\mu+\overline{\xi}_a\eta^a+\xi_a\overline{\eta}^a.
$$
On the other hand, the measure $DA_\mu^a D\eta^a D\overline{\eta}^a $ is BRST invariant, this is shown in pages 273-274 of \cite{ryder}. This implies that integration of BRST related configurations can be performed without changing the value of $G(J, \xi, \overline{\xi}, K, L)$. As the last lagrangian component $L_{nbrst}$  is not BRST invariant, and the overall result cannot change, this imposes constraints on the mean values  $<A^b_\mu>$, $<\eta^a>$, and $<\overline{\eta}^a>$  of the fields in $L_{nbrst}$. On the other hand, these mean values can be expressed using functional derivatives of $\Gamma(A, \eta, \overline{\eta}, K, L)$, as shown in \eqref{legendre}. For linear gauges, by making the redefinition $\Gamma\to \Gamma+\frac{\lambda}{2}\int d^4x F_a^2$ the resulting identity  is given by \cite{zinnjustin}-\cite{zinnjustin3}
\begin{equation}\label{slavnov}
\int d^4x\bigg[\frac{\delta \Gamma}{\delta  A_\mu^a }\frac{\delta \Gamma}{\delta K^\mu_a}+\frac{\delta  \Gamma}{\delta \eta^a}\frac{\delta \Gamma}{\delta L_a}\bigg]=0.
\end{equation}
Derivations can be found in pages 274-275 of \cite{ryder} or pages 597-599  of \cite{itzykson}. In addition to BRST symmetry, the other symmetry is the anti-ghost translation $\overline{\eta}^a\to \overline{\eta}^a+\delta\overline{\eta}^a$. This also is reflected in the mean values and at the end, results in the following constraint
\begin{equation}\label{ghostmotion}
\phi_\mu^{ab} \frac{\delta \Gamma}{\delta K_\mu^b}+\frac{\delta \Gamma}{\delta \overline{\eta}^a}=0.
\end{equation} 
This last equation is usually referred in the literature as the anti-ghost equation of motion, and leads to the functional dependence
\begin{equation}\label{funciona}
\Gamma(A, \eta, \overline{\eta}, K, L)=\Gamma(A, \eta, K_\mu^a-\overline{\eta}^b\phi^{ba}_\mu, L),
\end{equation}
of the effective action.  As it was remarked above, for general gauges the Slavnov-Taylor identities do not imply BRST invariance of $\Gamma(A, \eta, \overline{\eta}, K, L)$. In other words, the formula \eqref{generaltotal} may not apply for $\Gamma(A, \eta, \overline{\eta}, K, L)$. However, there is an important result on symmetries of the effective action which constraints it general form \cite{slavnov}-\cite{taylor}.
\\

\emph{Statement:} The effective action $\Gamma(A, \eta, \overline{\eta}, K, L)$ is invariant under a  symmetry of the original action $S_{gfs}(A, \eta, \overline{\eta}, K, L)$, if the action of this symmetry on the fields is \emph{linear}. 
\\

A proof can be found in pages 75-77 of \cite{weinberg} or in the survey \cite{bilal}.  The linear symmetries of the original action $S_{gfs}(A, \eta, \overline{\eta}, K, L)$ are ghost number conservation  and Lorentz transformations.  These symmetries are then expected to be preserved due to the above statement. However, this affirmation has to be taken with caution, as the renormalization process may break those symmetries. The dimensional regularization method \cite{bolo}-\cite{ash} is adequate for the preservation of those symmetries, and it is assumed on the following that this type of regularization is employed \cite{itzykson}-\cite{ryder}. With this caution in mind,  the action  $\Gamma(A, \eta, \overline{\eta}, K, L)$  has to have zero ghost number, it is invariant under Lorenz transformations, and has dimension four or less due to renormalizability. The only symmetry that is not linear in the fields is the BRST symmetry \eqref{s},  therefore, it cannot be assumed that $\Gamma(A, \eta, \overline{\eta}, K, L)$ is BRST invariant. The functional dependence \eqref{funciona} shows that the dependence on $K_\mu^a$ and $\overline{\eta}^a$ is through the combination   $K_\mu^a-\overline{\eta}^a\phi^{ab}_\mu$.  Therefore one has to consider the most general action depending on this combination, Lorentz invariant and with zero ghost number. This is precisely the situation considered in formulas \eqref{casigeneraltotal}-\eqref{generaltotaldes}. Therefore it may be concluded that
\begin{equation}\label{casigeneraltotalo}
\Gamma=\int  d^4 x[l(A^b_\mu)+(K^{\mu}_{a}-(\overline{\eta}\phi^\mu)_{a})\Delta_{\mu}^{ a b}\eta_{b}+\frac{1}{2}d_{abc}L_{a} \eta_{b} \eta_{c}],
\end{equation}
where now
$$
\Delta_{\mu}^{ a b}=\alpha \partial_\mu \delta_{ab}-\beta g c_{abc} A^c_{ \mu},\qquad d_{abc}=\gamma g c_{abc}.
$$
This is the second formula of page 602 of \cite{itzykson}. 

Some comments about this formula are mandatory. This formula will employed in the following chapters in order to study corrections to the initial action $S_{gfs}$, which is the gauge fixed Yang-Mills action. Even though in formulas \eqref{casigeneraltotal}-\eqref{generaltotaldes} the constants $d_{abc}$ were not related to the gauge group, it is expected that the corrections will not affect the gauge group of the theory, thus $d_{abc}$
are given by the gauge structure constants $c_{abc}$ up to a constant $\gamma$.  In addition, in deducing \eqref{generaltotaldes}  from \eqref{casigeneraltotal} a matrix $Q_{ab}$ was removed by a field redefinition, under the assumption that this matrix is invertible. As for the initial action  $S_{gfs}$ this matrix is $Q_{ab}\sim \delta_{ab}$, which is invertible, the general form \eqref{casigeneraltotalo} holds under the assumption that the corrections do not spoil this invertibility property. 

The formula \eqref{casigeneraltotalo} is pretty close to \eqref{generaltotal}. However, the difference is that the constants $\beta$ and $\gamma$ here are different in principle and that $\alpha$ is not necessarily equal to one. This difference appears since the BRST invariance and the consequent nilpotency conditions were not applied.  As before $l(A_\mu^a)$ has dimensions four or less and contains derivatives of the field, but it can not be assumed that is gauge invariant. It may contain terms of the form $\partial_\mu  A^a_\nu F^{a\mu\nu}$ or $A_\mu^a A^\mu_a$ and so on.  In fact, its explicit expression will be found in the next sections. 

 \subsection{Constraining the effective action by the Slavnov-Taylor identities}

Consider the expansion in $\hbar$ given above in \eqref{expansion}  for $\Gamma(A, \eta, \overline{\eta}, K, L)$, that is
$$
\Gamma=\Gamma_0+\hbar\Gamma_1+\hbar^2\Gamma_2+...
$$
 The zero term is identified with  the initial action
\begin{equation}\label{gut}
\Gamma_0=S_{(0)gf}(A_0, \eta_0, \overline{\eta}_0, K_0, L_0, g_0)=\int  d^4 x[L(A_0)+(K^{\mu}_{0a}-(\overline{\eta}_0\phi^\mu)_{a})D_{0\mu a b}\eta_{0b}+\frac{g}{2}c_{abc}L_{0a} \eta_{0b} \eta_{0c}],
\end{equation}
and the subscript "0" was included by convenience,  in order to indicate that this is the zero term of the recursive procedure  to be described below. In general every term in the expansion will contain a finite piece and a divergent one  $\Gamma_n=\Gamma^f_n+\Gamma_n^d$. The divergent part $\Gamma_n^d$ may be removed by adding counter terms to the action.  The Slavnov-Taylor identity, truncated to order $n$, is given by \cite{zinnjustin}-\cite{zinnjustin3}
$$
\sum_{p+q=n} \int d^4x\bigg[\frac{\delta \Gamma_p}{\delta  A^a_\mu }\frac{\delta \Gamma_q}{\delta K^\mu_a}+\frac{\delta  \Gamma_p}{\delta \eta^a}\frac{\delta \Gamma_q}{\delta L_a}\bigg]=0.
$$
In fact, to be more precise, this formula has to be true for all the lower orders, that is, for $p+q\leq n$. The zero action is $\Gamma_0(S_{(0)gf})=S_{(0)gf}$. This initial piece fulfills the Slavnov-Taylor identity
\begin{equation}\label{i0}
\int d^4x\bigg[\frac{\delta S_{(0)gf}}{\delta  A^a_\mu }\frac{\delta S_{(0)gf}}{\delta K_a^\mu}+\frac{\delta  S_{(0)gf}}{\delta \eta^a}\frac{\delta S_{(0)gf}}{\delta L_a}\bigg]=0.
\end{equation}
Consider now $\Gamma=\Gamma_0+\hbar \Gamma_1=S_{(0)gf}+\hbar \Gamma_1$. Then the Slavnov Taylor identity becomes
$$
\int d^4x\bigg[\frac{\delta S_{(0)gf}}{\delta  A^a_\mu}\frac{\delta \Gamma_1}{\delta K_a^\mu}+\frac{\delta  S_{(0)gf}}{\delta \eta^a}\frac{\delta \Gamma_1}{\delta L_a}+\frac{\delta \Gamma_1}{\delta  A^a_\mu }\frac{\delta S_{(0)gf}}{\delta K_a^\mu}+\frac{\delta  \Gamma_1}{\delta \eta^a}\frac{\delta S_{(0)gf}}{\delta L_a}\bigg]=0.
$$
This leads to an equation of the form
$$
\sigma_0\Gamma_1(S_{(0)gf})=\partial_\mu D^\mu,
$$
where the term $D^\mu$ inside the divergence has to have suitable vanishing conditions at the boundary of the space time, and  where the nilpotent operator 
$$
\sigma_0=\frac{\delta S_{(0)gf}}{\delta  A^{a}_{0\mu}}\frac{\delta }{\delta K_{0a}^\mu}+\frac{\delta  S_{(0)gf}}{\delta \eta_0^a}\frac{\delta }{\delta L_{0a}}+\frac{\delta S_{(0)gf}}{\delta  K_{0a}^\mu }\frac{\delta }{\delta A^a_{0\mu}}+\frac{\delta  S_{(0)gf}}{\delta L_{0a}}\frac{\delta }{\delta \eta_0^a},
$$
has been introduced.  Since $\Gamma_1$ has zero ghost number, is of dimension less than four, and is invariant under both the global ghost symmetry and Lorenz symmetry, its general form is given by \eqref{casigeneraltotalo}.  The resulting equation
\begin{equation}\label{sumo}
\frac{\delta S_{(0)gf}}{\delta  A^{a}_{0\mu}}\frac{\delta  \Gamma_1 }{\delta K_{0a}^\mu}+\frac{\delta  S_{(0)gf}}{\delta \eta_0^a}\frac{\delta  \Gamma_1 }{\delta L_{0a}}+\frac{\delta S_{(0)gf}}{\delta  K_{0a}^\mu }\frac{\delta  \Gamma_1}{\delta A^a_{0\mu}}+\frac{\delta  S_{(0)gf}}{\delta L_{0a}}\frac{\delta  \Gamma_1 }{\delta \eta_0^a}=\partial_\mu D^\mu.
\end{equation}
can be analyzed by use of the following formulas, which are derived diretly from \eqref{casigeneraltotalo} and \eqref{gut} 
$$
\frac{\delta  S_{(0)gf}}{\delta L_{0a}}=\frac{g}{2}c_{abc} \eta_{0b} \eta_{0c},\qquad \frac{\delta  \Gamma_1}{\delta L_{0a}}=\frac{1}{2}d_{abc} \eta_{0b} \eta_{0c},
$$
$$
\frac{\delta  S_{(0)gf}}{\delta \eta_{0a}}=-D_{0\mu}^{ a b}(K^{\mu}_{0b}-(\overline{\eta}_0\phi^\mu)_{b})+g c_{abc}  \eta_{0b} L_{0c},\qquad \frac{\delta  \Gamma_1}{\delta \eta_{0a}}=-\Delta_{1\mu}^{ a b}(K^{\mu}_{0b}-(\overline{\eta}_0\phi^\mu)_{b})+d_{abc}  \eta_{0b} L_{0c},
$$
$$
\frac{\delta  S_{(0)gf}}{\delta A_{0\mu}^{a}}=\frac{\delta  S_y}{\delta A_{0\mu}^{a}}+g c_{abc}(K^{\mu}_{0b}-(\overline{\eta}_0\phi^\mu)_{b})\eta_{0c},\qquad \frac{\delta  \Gamma_1}{\delta A^{a}_{0\mu}}=\frac{\delta  s}{\delta A_{0\mu}^{a}}+\beta_1 c_{abc}(K^{\mu}_{0b}-(\overline{\eta}_0\phi^\mu)_{b})\eta_{0c},
$$
\begin{equation}\label{totalidad}
\frac{\delta  S_{(0)gf}}{\delta K^\mu_{0a}}=D_{0\mu}^{ a b}\eta_{0b},\qquad \frac{\delta  \Gamma_1}{\delta K_{0a}^\mu}=\Delta_{1\mu}^{a b}\eta_{0b},
\end{equation}
where $S_y$ is the pure Yang-Mills action, $d_{abc}=\gamma_1 g c_{abc}$,
$$\Delta_{1\mu a b}=\alpha_1 \partial_\mu \delta_{ab}-\beta_1 g c_{abc} A^c_{ \mu},$$  and $s$ is given by
$$
s=\int l(A_0) d^4x,
$$
with $l(A)$ and $\alpha_1$, $\beta_1$ and $\gamma_1$ are defined in \eqref{casigeneraltotalo} and below. Recall that $l(A_0)$ has mass dimension four or less,  but it is not assumed to be gauge invariant. It has not to be so. In fact, the Slavnov-Taylor identities are usually employed to study the preservation of gauge invariance, so this can not be assumed from the beginning. Now, the type of terms corresponding to $S_{0gf}$ and $\Gamma_1$ have a similar structure, the main difference is that the constants $\alpha_1$, $\beta_1$ and $\gamma_1$ are turned on. The task is now to find the explicit solution of the equation \eqref{sumo}. In the following, two methods will be employed. One direct and laborious and other more sophisticated, which employs some specific properties of the operator $\sigma_0$.
\\

\emph{The straightforward method}
\\

The equation \eqref{sumo} can be solved by introducing the terms \eqref{totalidad} into it.  By applying this straightforward procedure, there appear terms with one, two and three ghost fields. The three ghost terms appear inside the expressions  $\frac{\delta  S_{(0)gf}}{\delta \eta_0^a}\frac{\delta  \Gamma_1 }{\delta L_{0a}}$ and $\frac{\delta \Gamma_1  }{\delta \eta_0^a}\frac{\delta  S_{(0)gf} }{\delta L_{0a}}$, and   vanish due to the Jacobi identity for $c_{abc}$. The other remaining  terms give rise  to two and one ghost term. The vanishing of these give information about the constants and the unknown functional $s(A_0)$ in \eqref{casigeneraltotalo}. After some integration by  parts if necessary, the vanishing of the two and one ghost terms leads to the identification $\beta_1=\gamma_1$ and
\begin{equation}\label{aboves}
D^\mu_{(0)ab}\frac{\delta s(A_0) }{\delta A^\mu_{0b}}+g(\beta_1-\alpha_1)c_{abc}A^\mu_{0c}\frac{\delta S_y(A_0) }{\delta A^\mu_{0b}}=0.
\end{equation}
Note that this formula is found exactly in page 602  of \cite{itzykson}. This equation should be employed in order to fix the unknown action $s(A_0)$ or equivalently, the lagrangian $l(A_0)$. 

As usual,  the full solution of a linear equation such as \eqref{aboves} can  be found by finding an homogeneous solution plus a particular one. The homogeneous equation is
$$
D^\mu_{(0)ab}\frac{\delta s_h(A_0) }{\delta A^\mu_{0b}}=0.
$$
This equation can be identified with the gauge invariant condition \eqref{gubki}, which leads to the conclusion that $s_h(A_0)$ is a gauge invariant functional. Furthermore, as $s_h(A_0)$  is Lorenz invariant and of dimension less or equal to four, it is clear that it is proportional to the Yang-Mills functional $S_y(A_0)$. Thus the general solution is
$$
s(A_0)=s_p(A_0)+a_1 S_y(A_0),
$$
with $a_1$ a new constant.  The task now is to find a particular solution $s_p(A_0)$ of equation \eqref{aboves}. This may not be easy, as there are several alternatives such as $\partial_\mu  A^a_\nu F_{\mu\nu}^a$ or $A_\mu^a A_\mu^a$, to cite some examples . However, some intuition can be gained by writing  explicitly the covariant derivative $ D_\mu^{ab}$ in \eqref{aboves},  a procedure that leads to the alternative expression
\begin{equation}\label{aboves2}
\partial_\mu\frac{\delta s_p(A_0) }{\delta A^\mu_{0a}}+g c_{abc}A^\mu_{0c}\frac{\delta s_p(A_0) }{\delta A^\mu_{0b}}+g(\beta_1-\alpha_1)c_{abc}A^\mu_{0c}\frac{\delta s_p(A_0) }{\delta A^\mu_{0b}}=0.
\end{equation}
The last two terms are similar, up to a $(\beta_1-\alpha_1)$ factor. This can be exploited in order to find the particular solution, provided this similarity is correctly interpreted. Recall that the Yang-Mills action is simply
\begin{equation}\label{heroe}
S_y(A_0)=-\frac{1}{4}\int d^4x F_{\mu\nu}^a F^{\mu\nu}_a,\qquad F_{\mu\nu}^a=\partial_\mu A^a_\nu-\partial_\nu A^a_\mu-gc_{abc}A^b_\mu A_\nu^c.
\end{equation}
The first two terms in the definition of $F_{\mu\nu}^a$ are homogeneous of first order in $A_\mu$, while the last one is homogeneous of order two. This in particular leads to the identity
$$
\frac{\delta F^a_{\mu\nu}}{\delta A_\alpha^b} A_\alpha^b=\partial_\mu A^a_\nu-\partial_\nu A^a_\mu-2gc_{abc}A^b_\mu A_\nu^c.
$$
The difference between the last expression $\frac{\delta F^a_{\mu\nu}}{\delta A_\alpha^b} A_\alpha^b$ and the definition of $F_{\mu\nu}^a$ is a factor of $2$ multiplying the  term proportional to $c_{abc}$. In other words, this term has been duplicated.  This situation resembles \eqref{aboves2}, where the last two terms look pretty similar. Besides, the mass dimensions of the fields $F_{\mu\nu}$ and $\frac{\delta F^a_{\mu\nu}}{\delta A_\alpha^b} A_\alpha^b$ are the same.  This perhaps non rigorous argument motivates to consider the trial action
\begin{equation}\label{trial2}
s_p=c\int d^4x \frac{\delta S_y}{\delta A_\alpha^b} A_\alpha^b,
\end{equation}
with $c$ an undetermined constant. The idea is to determine if there exists  a constant $c$ for which the last action is a solution of \eqref{aboves}, which is to be checked now. 

The first ingredient in order  to check that $s_p$ represents a solution is to derive the operator of equations of motion corresponding to the pure Yang-Mills theory. This is known to be
$$
\frac{\delta S_y}{\delta A_\mu^b}=D_\nu F^{a\nu\mu},
$$
 the equations of motion being identified with $D_\nu F^{a\nu\mu}=0$. From this, it follows that the above trial action \eqref{trial2} is, after an integration by parts,  given by
\begin{equation}\label{trial}
s_p=-c\int d^4x F^{a\mu\nu} D_\mu A_\nu^a=-\frac{c}{2}\int d^4x F^{a\mu\nu} (F^a_{\mu\nu}-gc_{abc}A^b_\mu A_\nu^c)=2c S_y+\Delta S.
\end{equation}
In other words, this action corresponds to a lagrangian which is the sum of the Yang-Mills one $2cL_y(A)$
and $2\Delta L=c gc_{abc}F_{\mu\nu}^a A^{b\mu} A^{c\nu}$. The Euler-Lagrange expressions resulting from the extra piece are of first order and reads as 
\begin{equation}\label{muevelo}
\frac{\delta \Delta S}{\delta A_\mu^a}=2c gc_{abc}[(\partial_\mu A^b_\nu)A^{c\nu }-2(\partial_\nu A^b_\mu)A^{c\nu}+(\partial_\nu A^{b\nu})A_\mu^c]-4c g^2 c_{abc}c_{bde}A^d_\mu A^e_\nu A^{c\nu}.
\end{equation}
This is a first order operator acting on the gauge field. In these terms, taking into account that the gauge invariance  \eqref{gubki} of the Yang-Mills action implies 
$$
D^\mu_{(0)ab}\frac{\delta S_y(A_0) }{\delta A^\mu_{0b}}=0,
$$
it follows that
\begin{equation}\label{aboves4}
D^\mu_{(0)ab}\frac{\delta s_p(A_0) }{\delta A^\mu_{0b}}=D^\mu_{(0)ab}\frac{\delta \Delta S(A_0) }{\delta A^\mu_{0b}}.
\end{equation}
As the equations of motion $\frac{\delta \Delta S}{\delta A_\mu^b}$ that follows from \eqref{muevelo} are of first order, it is seen that the quantity is of second order.  With the help of the last formulas, the equation \eqref{aboves} becomes
\begin{equation}\label{barry}
D^\mu_{(0)ab}\frac{\delta \Delta S(A_0) }{\delta A^\mu_{0b}}+g(\beta_1-\alpha_1)c_{abc}A^\mu_{0c}\frac{\delta S_y(A_0) }{\delta A^\mu_{0b}}=0.
\end{equation}
It is plausible that both quantities in the last equation cancel each other, as both are of second order, thus this order match. Even though this is a plausible argument, it remains to see if this equation is satisfied for a suitable choice of the constant $c$ in \eqref{trial2}. This may be realized or not, if this second case takes place it means that the trial ansatz \eqref{trial2} is not correct.  Fortunately, the last equation \eqref{barry} is satisfied by fixing $c=\beta_1-\alpha_1$.  The checking is a bit cumbersome, but direct. For instance, from \eqref{muevelo}
it follows to first order in $g$
$$
D^\mu_{(0)ab}\frac{\delta \Delta S(A_0) }{\delta A^\mu_{0b}}=2c gc_{abc}\partial^\mu[(\partial_\mu A^b_{0\nu})A_{0}^{c\nu}-2(\partial_\nu A^b_{0\mu})A_{0}^{c\nu}+(\partial_\nu A^{b\nu}_{0})A_{0\mu}^c]+O(g^2)
$$
$$
=2gc_{abc}[(\partial_\mu A^b_{0\nu})(\partial^\mu A_{0}^{c \nu})-2(\partial_\nu A^b_{0\mu})(\partial^\mu A_{0}^{c\nu})+(\partial_\nu A^{b\nu}_{0})(\partial_\mu A_{0}^{c\mu})+(\partial^\mu\partial_\mu A^b_{0\nu}) A_{0}^{c\nu}
$$
$$
-2(\partial^\mu\partial_\nu A^b_{0\mu})A_{0}^{c\nu}+(\partial^\mu\partial_\nu A^{b\nu}_{0})A_{0\mu}^{c}]+O(g^2).
$$
Due to the antisymmetry $c_{abc}=-c_{acb}$ it is seen that several terms vanish, and by rename of the indices it is easily seen that 
\begin{equation}\label{viel}
D^\mu_{(0)ab}\frac{\delta \Delta S(A_0) }{\delta A^\mu_{0b}}=-2c gc_{abc}\partial^\nu[\partial_\mu A^b_{0\nu}-\partial_\nu A^b_{0\mu}]A_{0}^{c\mu}+O(g^2).
\end{equation}
On the other hand from
$$
\frac{\delta S_y(A_0) }{\delta A^\mu_{0b}}=D_\nu F^{b\nu\mu}=\partial^\nu F^{b\nu\mu}-g c_{abc}F^{b\nu\mu} A_{0\nu}^c,
$$
it follows that
\begin{equation}\label{viel2}
g(\beta_1-\alpha_1)c_{abc}A^\mu_{0c}\frac{\delta S_y(A_0) }{\delta A^\mu_{0b}}=g(\beta_1-\alpha_1)c_{abc}\partial^\nu[\partial_\mu A^b_{0\nu}-\partial_\nu A^b_{0\mu}]A_0^{c\mu}+O(g^2).
\end{equation}
The two terms \eqref{viel}-\eqref{viel2} are equal if $c=\beta_|-\alpha_1$. Thus, with this choice, these terms satisfy \eqref{barry}.
The remaining terms of order $g^2$ or higher cancel  as well, although the checking is a bit more laborious and it is left for the reader.

The conclusion of this discussion is that the general solution of \eqref{aboves}
is given by
$$
s(A_0)=\int d^4x\bigg[a_1L(A_0)+(\beta_1-\alpha_1)A^{\mu }_{0a}\frac{\delta S(A_0) }{\delta A^\mu_{0a}}\bigg],\qquad L(A_0)=-\frac{1}{4}F_{\mu\nu}^a F^{\mu\nu}_a,
$$
and by taking into account  \eqref{casigeneraltotalo}, it is seen that the general effective action at this order is
$$
\Gamma_1=\int  d^4 x[a_1L(A_0)+(\beta_1-\alpha_1)A^{\mu }_{0a}\frac{\delta S(A_0) }{\delta A^\mu_{0a}}+\alpha_1 (K-\overline{\eta}\phi)^\mu_{0a}(D_{\mu}\eta)_{0a}
$$
\begin{equation}\label{bastantegeneral}
+g(\beta_1-\alpha_1) (K-\overline{\eta}\phi)^\mu_{0a}c_{abc} A_{0c\mu }\eta_{0b}+\frac{\beta_1 g}{2}c_{abc}L_{0a} \eta_{0b} \eta_{0c}].
\end{equation}
These are the formulas (12.164) of page 602 of \cite{itzykson}. Even though the obtained functional is not BRST invariant, its general form is important for deriving a  multiplicative renormalization procedure for Yang-Mills theories, which respects the BRST invariance order by order, as it will be exposed below. However, before dealing with this subject, it is instructive to derive \eqref{bastantegeneral} with a neater procedure.
\\

\emph{The sophisticated method}
\\

There is another more sophisticated proof for the solution \eqref{bastantegeneral} of the equation \eqref{sumo}.  The point is that the operator $\sigma$ defined in \eqref{sumo} is nilpotent of order 2, namely $\sigma^2=0$.  The proof is direct and can be found in page 601 of \cite{itzykson}. Then the solution of  \eqref{sumo} can be 
written as 
\begin{equation}\label{potonto}
\Gamma_1=a_1\int  L(A_0) d^4x+\sigma X( (K-\overline{\eta}\phi)^\mu_{0a}, A_{0\mu}^a, L_{0a})
\end{equation}
where $X$ is a generic functional of the fields  $(K-\overline{\eta}\phi)^\mu_{0a}$, $\eta^a$, $A_{0\mu}^a$ and $L_{0a}$. This cohomological method was conjectured in \cite{kluberg} and proved in \cite{joglekar}. At first sight, the functional $X( (K-\overline{\eta}\phi)^\mu_{0a}, A_{0\mu}^a, L_{0a})$ can be anything. However,  it is direct from the definition of the operator $\sigma$, quoted below for convenience
\begin{equation}\label{mentira}
\sigma_0X=\frac{\delta S_{(0)gf}}{\delta  A^{a}_{0\mu}}\frac{\delta X}{\delta K_{0a}^\mu}+\frac{\delta  S_{(0)gf}}{\delta \eta_0^a}\frac{\delta X}{\delta L_{0a}}+\frac{\delta S_{(0)gf}}{\delta  K_{0a}^\mu }\frac{\delta X}{\delta A^a_{0\mu}}+\frac{\delta  S_{(0)gf}}{\delta L_{0a}}\frac{\delta X}{\delta \eta_0^a},
\end{equation}
that its action increases the ghost number and the mass dimensions of $X( (K-\overline{\eta}\phi)^\mu_{0a}, A_{0\mu}^a, L_{0a})$ by one. The action also respects Lorenz invariance. Therefore the task is to construct a functional  $X( (K-\overline{\eta}\phi)^\mu_{0a}, A_{0\mu}^a, L_{0a})$ with ghost number -1, mass dimension three and Lorenz invariant. It is not difficult to see that the candidate is 
$$
X= cL_{0a} \eta_{0a}+d(K-\overline{\eta}\phi)^\mu_{0a} A^a_{0\mu},
$$
where $c$ and $d$ are constants. It is direct to see that
$$
\frac{\delta X}{\delta  A^{a}_{0\mu}}=d(K-\overline{\eta}\phi)^\mu_{0a},\qquad \frac{\delta X}{\delta K_{0\mu}^a}=cA_{0\mu}^a,\qquad \frac{\delta X}{\delta  L_{0a}}=c\eta_{0a},\qquad \frac{\delta X}{\delta  \eta_{0a}}=cL_{0a},
$$
and this, together with \eqref{totalidad} leads after some direct calculation for  $\sigma_0 X$ given in \eqref{mentira} and to conclude that
$$
\Gamma_1=\int  d^4 x[a_1L(A_0)+d A^{\mu }_{0a}\frac{\delta S(A_0) }{\delta A^\mu_{0a}}-(c+d) (K-\overline{\eta}\phi)^\mu_{0a}(D_{\mu}\eta)_{0a}
$$
\begin{equation}\label{alucina}
+g d (K-\overline{\eta}\phi)^\mu_{0a}c_{ a b c} A_{0c\mu }\eta_{0b}-\frac{c g}{2}c_{abc}L_{0a} \eta_{0b} \eta_{0c}].
\end{equation}
The last action is the same as the non gauge invariant part of \eqref{bastantegeneral} if $c=-\beta_1$ and $d=\beta_1-\alpha_1$. In other words, the general formula \eqref{bastantegeneral} is recovered with this approach. In making this checking care has to be taken by the anti commuting nature of some of the fields. This procedure is much less cumbersome that the one described in \eqref{barry}, and this is the advantage of the above cohomological method.
 
\subsection{Multiplicative renormalization}
The Yang-Mills Lagrangian $L(A_0)=-\frac{1}{4}
F^a_{\mu\nu}F^{a\mu\nu}$ has the homogeneity property
$$
2L(A_0)=A^{\mu }_{0a}\frac{\delta S_y(A_0) }{\delta A^\mu_{0a}}-g_0\frac{\delta S_y(A_0) }{\delta g_0}.
$$
This property follows directly from the definition $F_{\mu\nu}^a=\partial_\mu A^a_\nu-\partial_\nu A^a_\mu-gc_{abc}A^b_\mu A_\nu^c$, which implies that the terms quadratic in $A_{0\mu}^a$  or its derivatives in the Yang-Mills lagrangian do not involve $g_0$, the cubic are proportional to $g_0$ and the quartic to $g_0^2$, whence the result. By extracting $A^{\mu }_{0a}\frac{\delta S_y(A_0) }{\delta A^\mu_{0a}}$ from the last expression and by inserting the result into \eqref{bastantegeneral} it is concluded that
$$
\Gamma_1=\int  d^4 x[(\beta_1-\alpha_1+\frac{a_1}{2})A^{\mu }_{0a}\frac{\delta S_y(A_0) }{\delta A^\mu_{0a}}-\frac{a_1g_0}{2}\frac{\delta S_y(A_0) }{\delta g_0}+\alpha_1 (K-\overline{\eta}\phi)^\mu_{0a}(D_{\mu}\eta)_{0a}
$$
\begin{equation}\label{sanatorio}
+g(\beta_1-\alpha_1) (K-\overline{\eta}\phi)^\mu_{0a}C_{ a bc} A_{0c\mu }\eta_{0b}+\frac{\beta_1 g}{2}C_{abc}L_{0a} \eta_{0b} \eta_{0c}].
\end{equation}
The expression above is written in terms of functional derivatives of the Yang-Mills  action and further terms involving the fields of the model. One of the key facts in proving multiplicative renormalization is that the last expression for $\Gamma_1$ can be expressed entirely in terms of functional derivatives of the action $S_{0gf}$.
In order to prove this,  a first step is to deduce from \eqref{gut}  the following identities
$$
K^{\mu}_{0a}\frac{\delta S_{0gf}}{\delta K^{\mu}_{0a}}=K^{\mu}_{0a} D_{0\mu a b}\eta_{0b},\qquad \overline{\eta}_{0a}\frac{\delta S_{0gf}}{\delta \overline{\eta}_{0a}}=-(\overline{\eta}_0\phi^\mu)_{a}D_{0\mu a b}\eta_{0b},\qquad  L_{0a}\frac{\delta S_{0gf}}{\delta L_{0a}}=\frac{g}{2}c_{abc}L_{0a} \eta_{0b} \eta_{0c},
$$
$$
\eta_{0a}\frac{\delta S_{0gf}}{\delta \eta_{0a}}=(K^{\mu}_{0a}-(\overline{\eta}_0\phi^\mu)_{a})D_{0\mu a b}\eta_{0b}+gc_{abc}L_{0a} \eta_{0b} \eta_{0c},$$
$$
g\frac{\delta S_{0gf}}{\delta g}=g\frac{\delta S_y(A_0)}{\delta g}+(K^{\mu}_{0a}-(\overline{\eta}_0\phi^\mu)_{a})gc_{a bc}\eta_{0b}A_{0c}^\mu+\frac{g}{2} c_{abc}L_{0a} \eta_{0b} \eta_{0c},
$$
$$
A^{\mu}_{0a}\frac{\delta S_{0gf}}{\delta A^{\mu}_{0a}}=A^{\mu}_{0a}\frac{\delta S_y(A_0)}{\delta A^{\mu}_{0a}}+(K^{\mu}_{0a}-(\overline{\eta}_0\phi^\mu)_{a})gc_{a bc}\eta_{0b}A_{0c}^\mu.
$$
This shows that
$$
K^{\mu}_{0a}\frac{\delta S_{0gf}}{\delta K^{\mu}_{0a}}+\overline{\eta}_{0a}\frac{\delta S_{0gf}}{\delta \overline{\eta}_{0a}}=(K^{\mu}_{0a}-(\overline{\eta}_0\phi^\mu)_{a})D_{0\mu a b}\eta_{0b},
$$
$$
\eta_{0a}\frac{\delta S_{0gf}}{\delta \eta_{0a}}-2 L_{0a}\frac{\delta S_{0gf}}{\delta L_{0a}}=(K^{\mu}_{0a}-(\overline{\eta}_0\phi^\mu)_{a})D_{0\mu a b}\eta_{0b},
$$
from which follows that linear dependence relation between the functional derivatives
\begin{equation}\label{dependencia}
K^{\mu}_{0a}\frac{\delta S_{0gf}}{\delta K^{\mu}_{0a}}+\overline{\eta}_{0a}\frac{\delta S_{0gf}}{\delta \overline{\eta}_{0a}}-\eta_{0a}\frac{\delta S_{0gf}}{\delta \eta_{0a}}=2 L_{0a}\frac{\delta S_{0gf}}{\delta L_{0a}}.
\end{equation}
Therefore, these derivatives are not all independent on each other.  In addition 
$$
A^{\mu}_{0a}\frac{\delta S_{0gf}}{\delta A^{\mu}_{0a}}-g\frac{\delta S_{0gf}}{\delta g}=A^{\mu}_{0a}\frac{\delta S_y(A_0)}{\delta A^\mu_{0a}}-g\frac{\delta S_y(A_0)}{\delta g}+L_{0a}\frac{\delta S_{0gf}}{\delta L_{0a}}.
$$
This leads to
$$
\hbar\Gamma_1=\int  d^4 x[(\beta_1-\alpha_1+\frac{a_1}{2})A^{\mu }_{0a}\frac{\delta S_{0gf} }{\delta A^\mu_{0a}}-\frac{a_1 g_0}{2}\frac{\delta S_{0gf} }{\delta g_0}+\alpha_1\bigg( K^{\mu}_{0a}\frac{\delta S_{0gf}}{\delta K^{\mu}_{0a}}+\overline{\eta}_{0a}\frac{\delta S_{0gf}}{\delta \overline{\eta}_{0a}}\bigg)
$$
$$
+(\frac{a_1}{2}+\beta_1)L_{0a}\frac{\delta S_{0gf}}{\delta L_{0a}}].
$$
This expression is not unique, since due to the linear dependence property \eqref{dependencia}, it may be written as
$$
\hbar\Gamma_1=\int  d^4 x\bigg[\bigg(\beta_1-\alpha_1+\frac{a_1}{2}\bigg) \bigg(A^{\mu }_{0a}\frac{\delta  }{\delta  A^\mu_{0a}}+L_{0a }\frac{\delta }{\delta L_{0a}}\bigg)
$$
\begin{equation}\label{sanatorio2}
+\frac{\alpha_1}{2}\bigg(K_{0a}\frac{\delta }{\delta K_{0a}}+\eta_{0a}\frac{\delta }{\delta \eta_{0a}}+\overline{\eta}_{0a}\frac{\delta }{\delta\overline{\eta}_{0a}}\bigg)-\frac{a_1g_0}{2} \frac{\delta }{\delta g_0}\bigg]S_{0gf}(A_0, \eta_0, \overline{\eta}_0, K_0, L_0, g_0).
\end{equation}
This is the formula (12.165) of page 602 of \cite{itzykson} and is fundamental for proving multiplicative renormalization \cite{dixon}. The last expression can be though as the analog  of  a tangent plane approximation, generalized for the functional case. It describes $\hbar \Gamma_1$ as the variation of $S_{0gf}$ along the "point" $(A_0, \eta_0, \overline{\eta}_0, K_0, L_0, g_0)$ by some increment depending on $\alpha_1$, $\beta_1$, and $a_1$. Explicitly, after a redefinition $\Gamma_1\to -\Gamma_1$ it follows that this incremental expression is
$$
\hbar\Gamma_1(S_{0gf})=-S_{0gf}(A_0, \eta_0, \overline{\eta}_0, K_0, L_0, g_0)
$$
\begin{equation}\label{reachingol}
+S_{0gf}\bigg((1+Z_1^A) A_0, (1+ Z_1^\eta) \eta_0, Z_1^\eta\overline{\eta}_0, (1+Z_1^\eta) K_0, (1+Z_1^A) L_0, (1+Z_1^g) g_0\bigg)
+O(\hbar^2),
\end{equation}
where
\begin{equation}\label{muzhina}
Z^A_1=\beta_1-\alpha_1+\frac{a_1}{2},\qquad Z_1^\eta=\frac{\alpha_1}{2},\qquad Z_1^g=\frac{a_1}{2}.
\end{equation}
The quantities $\beta_1$, $\alpha_1$ can be considered are first order in $\hbar$, and at this order the last expression  may be written, at the same order, as
\begin{equation}\label{reaching}
\hbar\Gamma_1(S_{0gf})=S_{0gf}\bigg(Z_1^A A_0, Z_1^\eta \eta_0, Z_1^\eta\overline{\eta}_0, Z_1^\eta K_0, Z_1^A L_0, Z_1^g g_0\bigg)
+O(\hbar^2).
\end{equation}
On the other hand, the first order term $\Gamma_1(S_{0gf})$ can be decomposed into a finite piece and a divergent one $\Gamma_1(S_{0gf})=\Gamma^f_1(S_{0gf})+\Gamma^d_1(S_{0gf})$. That is
\begin{equation}\label{reaching2}
\hbar\Gamma^d_1(S_{0gf})=S_{0gf}\bigg(Z_{1d}^A A_0, Z_{1d}^\eta \eta_0, Z_{1d}^\eta\overline{\eta}_0, Z_{1d}^\eta K_0, Z_{1d}^A L_0, Z_{1d}^g g_0\bigg)
+O(\hbar^2),
\end{equation}
\begin{equation}\label{reaching3}
\hbar\Gamma^f_1(S_{0gf})=S_{0gf}\bigg(Z_{1f}^A A_0, Z_{1f}^\eta \eta_0, Z_{1f}^\eta\overline{\eta}_0, Z_{1f}^\eta K_0, Z_{1f}^A L_0, Z_{1f}^g g_0\bigg)
+O(\hbar^2),
\end{equation}
where the constants with the subindex $f$ and $d$ are such that their sum is \eqref{muzhina}.
Therefore,
\begin{equation}\label{momento}
\Gamma(S_{0gf})=S_{0gf}+\hbar \Gamma^f_1(S_{0gf})+\hbar \Gamma^d_1(S_{0gf})+O(\hbar^2).
\end{equation}
The first order effective action \eqref{reaching} was determined by the condition $\sigma \Gamma_1=0$, however, both the divergent part and the finite one have to satisfy this equation independently
\begin{equation}\label{both}
\sigma \Gamma_1^f=0,\qquad \sigma \Gamma_1^d=0,
\end{equation}
as they can not cancel each other. The divergent part in \eqref{momento} can be deleted by the redefinition 
\begin{equation}\label{aimitar}
S_{0gf}\to S_{1gf}=S_{0gf}-\hbar \Gamma_1^d(S_{0gf}),
\end{equation}
which, by use of \eqref{reaching3} can be written as
\begin{equation}\label{conclu}
S_{1gf}=S_{0gf}\bigg((1-Z_{1d}^A) A_0, (1-Z_{1d}^\eta) \eta_0,(1- Z_{1d}^\eta)\overline{\eta}_0, (1-Z_{1d}^\eta) K_0, (1-Z_{1d}^A) L_0, (1-Z_{1d}^g) g_0\bigg)
+O(\hbar^2).
\end{equation}
The problem now is that the following identification $\Gamma_0=S_{1gf}$ holds. This means that $S_{1gf}$ must satisfy the analogous of equation \eqref{i0}, namely
\begin{equation}\label{disco}
\int d^4x\bigg[\frac{\delta S_{1gf}}{\delta  A^a_\mu }\frac{\delta S_{1gf}}{\delta K_a^\mu}+\frac{\delta  S_{1gf}}{\delta \eta^a}\frac{\delta S_{1gf}}{\delta L_a}\bigg]=0.
\end{equation}
 It can be directly checked that the equation \eqref{disco}, which insures BRST invariance for $S_{1gf}$ is not satisfied. Instead, it is given by
\begin{equation}\label{discu}
\int d^4x\bigg[\frac{\delta S_{1gf}}{\delta  A^a_\mu }\frac{\delta S_{1gf}}{\delta K_a^\mu}+\frac{\delta  S_{1gf}}{\delta \eta^a}\frac{\delta S_{1gf}}{\delta L_a}\bigg]=\hbar^2\int d^4x\bigg[\frac{\delta \Gamma^d_1}{\delta  A^a_\mu }\frac{\delta \Gamma^d_1}{\delta K_a^\mu}+\frac{\delta  \Gamma^d_1}{\delta \eta^a}\frac{\delta \Gamma^d_1}{\delta L_a}\bigg],
\end{equation}
where the equations \eqref{both} were taken into account. The conclusion is that the required condition is violated. However, it is violated at order $\hbar^2$; thus, at first order it is satisfied. The exact identification is  
$S_{1gf}=S_{0gf}-\hbar \Gamma_1^d(S_{0gf})+\Delta$ where the last part $\Delta$ may be functionally complicated but only affect higher orders, as it is of order $O(\hbar^2)$. At linear order $\Delta$ may be omitted.

The expression \eqref{conclu} has the same functional form as the original action but with rescaled fields. This rescaling is necessary for handling divergences. By taking into account formula \eqref{momento}, this leads to
\begin{equation}\label{renormal}
\Gamma^{ren}=S_{0gf}\bigg((1-Z_{1d}^A) A_0, (1-Z_{1d}^\eta) \eta_0,(1- Z_{1d}^\eta)\overline{\eta}_0, (1-Z_{1d}^\eta) K_0, (1-Z_{1d}^A) L_0, (1-Z_{1d}^g) g_0\bigg)+\hbar \Gamma_1^f.
\end{equation}
By use of \eqref{reaching3} it is finally arrived to
\begin{equation}\label{renormal2}
\Gamma^{ren}=S_{1gf}(A_1, \eta_1, \overline{\eta}_1, K_1, L_1, g_1)+O(\hbar^2)
\end{equation}
$$
=S_{0gf}\bigg((1-Z_{1}^A) A_0, (1-Z_{1}^\eta) \eta_0, (1-Z_{1}^\eta)\overline{\eta}_0, (1-Z_{1}^\eta) K_0, (1-Z_{1}^A) L_0,(1- Z_{1}^g) g_0\bigg)+O(\hbar^2).
$$
Here the divergent constants disappeared and now
\begin{equation}\label{constantinopla}
Z^A_{1}=1+\beta_f-\alpha_f+\frac{a_f}{2},\qquad Z_{1}^\eta=1+\frac{\alpha_f}{2},\qquad Z_{1}^g=1+\frac{a_f}{2}.
\end{equation}
As the renormalized effective action \eqref{renormal2} is given in terms of the original action $S_{0gf}$, it is seen that after a redefinition of the fields  that the action is approximately BRST invariant at first order, as the original action is invariant under this symmetry.  This result is true up to higher order terms. The renormalization is achieved by multiplicative redefinition of the fields of the model and coupling constants, as seen from the last two formulas.

The procedure above from zero to one order can be generalized by taking $S_{1gf}$ in \eqref{renormal2}
and repeating the procedure for finding $\Gamma^{ren}$ to second order. Again, multiplicative renormalization holds, since the procedure is completely analogous as the first order one. This procedure can be applied order by order, resulting in
\begin{equation}\label{renormaln}
\Gamma^{ren}=S_{(n)gf}(A_n, \eta_n, \overline{\eta}_n, K_n, L_n, g_n)
\end{equation}
$$
=S_{(n-1)gf}\bigg((1-Z_{n}^A) A_{n-1}, (1-Z_{n}^\eta) \eta_{n-1}, (1-Z_{n}^\eta)\overline{\eta}_{n-1}, (1-Z_{n}^\eta) K_{n-1}, (1-Z_{n}^A) L_{n-1}, (1- Z_{n}^g)g_{n-1}\bigg),
$$
up to terms of order $O(\hbar^{n+1})$. This procedure describe the $n$-order term and fields in terms of the $(n-1)$ order ones times some normalization constants. Again the result is given in terms of $S_{ngf}$ which is inductively BRST invariant

In brief, the discussion given above shows that the effective action is renormalized by multiplying the fields by certain constants. These multiplication constants are not all independent \eqref{constantinopla}, some of them are identical to others as a consequence of the Slavnov-Taylor identities. All linear symmetries of the original action are maintained as symmetries of the effective actions. Even though the BRST symmetry is not exactly preserved, it remains valid order by order, up to higher-order terms in $\hbar$. 

The next step is to repeat this analysis for gravity theories. The methods developed above will be useful for these matters.

\section{Quadratic Quantum Gravity}

\subsection{Diffeomorphisms in gravity theories}
Consider a theory of pure gravity,  whose relevant degree of freedom is the metric tensor $g_{\mu\nu}$ defining the Lorenz interval
$$
ds^2=g_{\mu\mu}(y)dy^\mu dy^\nu,
$$
in a curved space $M_4$. The classical equations of motion for the model are assumed to be covariant under generic coordinate transformations. The most studied example of such type of theories is GR. Under general coordinate transformations $y^\mu\to x^\mu(y^\nu)$, the metric tensor $g_{\mu\nu}$ is replaced by $\widetilde{g}_{\mu\nu}$
$$
\widetilde{g}_{\alpha\beta}(x)=\frac{\partial y^\mu}{\partial x^\alpha} \frac{\partial y^\nu}{\partial x^\beta}g_{\mu\nu}(y).
$$
For small diffeomorphisms  $y^\nu= x^\mu+\xi^\mu(x)$, where $\xi^\mu$ are infinitesimal quantities, it follows that $\partial_\alpha y^\mu=\delta^{\mu}_\alpha+\partial_\alpha \xi^\mu$ and this, together a Taylor expansion of $g_{\mu\nu}(y)$ around the point $x^\mu$ leads to an infinitesimal variation of the metric
\begin{equation}\label{gaugegra}
\delta g_{\mu\nu}=\xi^\alpha \partial_\alpha g_{\mu\nu}+g_{\mu\alpha}\partial_\nu \xi^\alpha+g_{\nu\alpha}\partial_\mu \xi^\alpha,\end{equation}
up to a second order terms. Here $\delta g_{\mu\nu}=\widetilde{g}_{\mu\nu}(x)-g_{\mu\nu}(x)$ is the variation of the metric at the same point $x^\mu$.

Diffeomorphisms in gravity theories lead to problems similar to those in Yang-Mills theories.  Two metric tensors $g_{\mu\nu}$ and  $g^\xi_{\mu\nu}$ are, loosely speaking,  equivalent if they are related by a diffeomorphism. This affirmation has to be taken with care, as a change of coordinates may enlarge or restrict the domain where the space time is defined, and there can be locally equivalent metrics which are defined on topologically different space times.  In any case, the quantization of gravity theories by the method of path integrals has an over counting problem, since the functional integration on the space of possible metric tensors is made over a given tensor $g_{\mu\nu}$ and all its equivalent representatives $g^\xi_{\mu\nu}$ under the action of diffeomorphisms $\xi$, which are generically an infinite set.  This problem has to be cured by a Faddev-Popov procedure.

It is customary to introduce the metric perturbation $\widetilde{h}_{\mu\nu}$ given by $g_{\mu\nu}=\eta_{\mu\nu}+\kappa \widetilde{h}_{\mu\nu}$ where $\kappa=\sqrt{16\pi G_N}$ is the GR gravitational coupling. Under infinitesimal changes of coordinates, by use of \eqref{gaugegra} it is obtained directly that
\begin{equation}\label{gaugegra2}
\delta  \widetilde{h}_{\mu\nu}=\xi^\alpha \partial_\alpha  \widetilde{h}_{\mu\nu}+ \widetilde{h}_{\mu\alpha}\partial_\nu \xi^\alpha+ \widetilde{h}_{\nu\alpha}\partial_\mu \xi^\alpha+\frac{1}{\kappa}\eta_{\mu\alpha}\partial_\nu \xi^\alpha+\frac{1}{\kappa}\eta_{\nu\alpha}\partial_\mu \xi^\alpha.
\end{equation}
In some applications it is usual to introduce another perturbation, denoted by $h^{\mu\nu}$ and defined by
\begin{equation}\label{pertu}
\kappa h^{\mu\nu}=\sqrt{-g}g^{\mu\nu}-\eta^{\mu\nu}.
\end{equation}
The change under infinitesimal diffeomorphisms for $h^{\mu\nu}$ can be found by use of \eqref{gaugegra2} by the following procedure. First employ the formula
$$
\sqrt{-g}=\sqrt{\det(-\eta-\kappa \widetilde{h})}=\exp{\frac{1}{2}\log(\det(-\eta-\kappa \widetilde{h}))}=\exp{\frac{1}{2}\log(\det(-1-\kappa\eta \widetilde{h}))},
$$
the last step employs that $\eta_{\mu\nu}=\eta^{-1}_{\mu\nu}$. By taking into account the identity "$\log \det=\text{Tr}\log$" it follows that
$$
\sqrt{-g}=\exp{\frac{1}{2}\text{Tr}(\log(-1-\kappa\eta \widetilde{h}))},
$$
and by a simple Taylor expansion of the exponential  and the logarithm it is found that
$$
\sqrt{-g}\simeq 1+\frac{1}{2} \widetilde{h}_\alpha^\alpha+\frac{1}{8} \widetilde{h}_\alpha^\alpha \widetilde{h}_\beta^\beta-\frac{1}{4} \widetilde{h}_\alpha^\beta  \widetilde{h}_\beta^\alpha.
$$
The definition \eqref{pertu} then leads to 
\begin{equation}\label{pertu2}
\kappa h^{\mu\nu}=\frac{1}{2}\eta^{\mu\nu} \widetilde{h}_\alpha^\alpha-\widetilde{h}^{\mu\nu}.
\end{equation}
The last expression in addition, implies that $\kappa h_\alpha^\alpha=\widetilde{h}_\alpha^\alpha$ and this leads to the inversion formula
\begin{equation}\label{pertu3}
\widetilde{h}^{\mu\nu}=\frac{\kappa}{2}\eta^{\mu\nu}h_\alpha^\alpha-\kappa h^{\mu\nu}.
\end{equation}
The transformation of $h^{\mu\nu}$ then follows from \eqref{pertu2} and \eqref{gaugegra2}, together with \eqref{pertu3}.
The result is
\begin{equation}\label{infdif}
\delta_\xi h^{\mu\nu}=D_\alpha^{\mu\nu}\xi^\alpha= \partial^\mu \xi^\nu +\partial^\nu \xi^\mu-\eta^{\mu\nu}\partial_\alpha \xi^\alpha
+ h^{\alpha\mu} \partial_\alpha \xi^\nu+ h^{\alpha\nu} \partial_\alpha \xi^\mu-\xi^\alpha\partial_\alpha h^{\mu\nu}-h^{\nu\nu}\partial_\alpha \xi^\alpha.
\end{equation}
The new operator $D_\alpha^{\mu\nu}$ is simply defined by the last identity. In the following, $h^{\mu\nu}$ is the degree of freedom to be considered, and the action of the infinitesimal diffeomorphism operator $D_\alpha^{\mu\nu}$ will play the analogous role of a gauge transformation in Yang-Mills theories.

\subsection{The non renormalizability of GR and renormalizable generalizations}
Even taking into account the above analogy with Yang-Mills theories, a fundamental difference is  that GR is not renormalizable. Before giving an expanded explanation, a  brief non renormalizability argument can be presented, that goes as follows \cite{feynman}, \cite{dewitt3}. The Einstein-Hilbert action of GR is well known 
\begin{equation}\label{GR}
S_{GR}=\frac{1}{16\pi G_N}\int \sqrt{-g}\; R\;d^4x. 
\end{equation}
Here $R(g_{\mu\nu})$ is the curvature scalar.  Recall that the Riemann curvature tensor, in a local system of coordinates, can be calculated in terms of the Christofel symbols 
$$
\Gamma_{\mu\nu}^{\alpha}=\frac{1}{2}g^{\alpha\beta}(\partial_{\mu} g_{\beta\nu}+\partial_{\nu}g_{\beta\mu}-\partial_{\beta}g_{\mu\nu}),
$$
as follows
$$
R^\beta_{\alpha \mu\nu}=\partial_\mu \Gamma_{\nu\alpha}^\beta-\partial_\nu \Gamma_{\mu\alpha}^\beta+\Gamma_{\mu\gamma}^\beta\Gamma_{\nu\alpha}^\gamma-\Gamma_{\nu\gamma}^\beta\Gamma_{\mu\alpha}^\gamma.
$$
This tensor may be expanded order by order in $h^{\mu\nu}$ by use of formulas such as \eqref{pertu}-\eqref{infdif}  and higher order analogs. For instance
$$
\Gamma_{\mu\nu}^{\alpha}=\frac{1}{2}\eta^{\alpha\beta}(\partial_{\mu} h_{\beta\nu}+\partial_{\nu}h_{\beta\mu}-\partial_{\beta}h_{\mu\nu})+O(h^2),
$$
$$
R^\alpha_{\rho\mu\nu}=\frac{1}{2}\eta^{\alpha\beta}(\partial_\rho\partial_{\mu}h_{\beta\nu}-\partial_\rho\partial_{\nu}h_{\beta\mu}-\partial_\nu\partial_{\beta}h_{\rho\mu}+\partial_\mu\partial_{\beta}h_{\rho\nu})+O(h^2),
$$
and $R_{\mu\nu}=\eta^{\alpha\beta}R_{\mu\alpha\nu\beta}$, $R=\eta^{\alpha\beta}R_{\alpha\beta}$ up to higher orders in the perturbation $h^{\mu\nu}$. 

The expansion in $h^{\mu\nu}$ leads to non interacting terms, plus interacting ones from where the interaction vertices can be read. The non interacting terms are collected in the so called zero order action, which has a generic form
\begin{equation}\label{propaga}
S_{(0)GR}=\int  h^{\mu\nu}D^{-1}_{\mu\nu,\alpha\beta} h^{\alpha\beta} d^4x.
\end{equation}
Here $D_{\mu\nu, \alpha\beta}$ is the standard propagator, and the last zeroth order action is written in terms of its inverse. The propagator in standard GR contains second order derivatives.  This implies that the quantity $D^{-1}_{\mu\nu,\alpha\beta}$, in momentum space and for GR, goes as $k^{-2}$ after appropriate gauge fixing. In addition, the higher order terms give rise to $n$-graviton vertices, with $n$ being any possible integer. This is different for Yang-Mills theories, where there are only three and four vertex interactions for the gauge fields. 

Another difference between gravity and Yang-Mills theories is that the gauge coupling constant $g$ of the former is dimensionless while gravity contains the Planck mass $M_p$ present in the Newton constant $G_N$. In momentum space, the $n$-vertex interactions may be complicated expressions in the incoming and outgoing momenta \cite{prinz}. Despite this complexity, the resulting expression is quadratic in momentum, this is intuitive as the GR action is dimensionless and the square of the Planck mass $M^2_p$ is a common factor, thus only a further quadratic mass scale is needed, which is given by the momenta $p_i$ of the particles participating in the vertex interaction.  With these details in mind, consider a Feynman graph, with  $V_n$ vertices, with $n$ the number of gravitons incoming or outgoing from it. This graph has generic $L_i$ internal lines and $K$ the number of independent momentum integrations. These quantities are not independent, they are related by the topological relation for Feynmann diagrams
 \begin{equation}\label{sobre}
K=L_i-\sum V_n+1.
\end{equation}
The superficial degree of divergence of a given  diagrams containing gravitons is then
given by
$$
D=-2L_i+2\sum_n V_n+4K.
$$
The factor $2$ multiplying $L_i$ indicates that the propagator goes as $k^{-2}$ and the factor $2$ in $V_n$ indicates that the vertices are quadratic in momentum. The factor $4$ multiplying $K$ is due to the fact that the theory is defined in four dimensions, thus the momenta has four components. The combination of the last two formulas leads to
$$
D=-2(K-1)+4K=2(K+1).
$$
This divergence grows as the number $K$ of momentum integrations increases, which makes the renormalizability of the theory compromised.

The argument given above about the non renormalizability may be avoided  if GR  is generalized to  a theory for which the degree of divergence is of the form
$$
D_n=-4L_i+4\sum_n V_n+4K,
$$
which, by use of \eqref{sobre}, is given by
\begin{equation}\label{nocrece}
D_n=-4(K-1)+4K=4.
\end{equation}
This divergence degree for this hypothetic new theory does not grow with $K$, which suggests that such theory may be renormalizable. The difference between $D$ and $D_n$ is that the factor $2$ multiplying the internal line and vertex numbers has been replaced by a factor $4$. In other words, this theory has to have a propagator that goes as $k^{-4}$ instead of $k^{-2}$ and the vertices are quartic in momentum instead of quadratic \cite{dewitt3}.  

Theories with propagators which goes as $k^{-4}$ for large $k$ are known in the literature. Examples are Ostrogradskyi oscillators with fourth order equations of motion \cite{pais}.  Another example  appears naturally in the  Pauli-Villars renormalization method for scalar fields in QFT. Given for instance a massive scalar, this method renormalize certain divergent diagrams with the replacement of the propagator
$$
D_\phi=\frac{1}{k^2+m^2}\to D_\phi=\frac{1}{k^2+m^2}-\frac{1}{k^2+M^2}\sim \frac{1}{k^4},
$$
the last behavior is valid for large $k$. The Pauli-Villars procedure involves the introduction of a mass scale $M$ corresponding to a new scalar particle. This modification allows to renormalize the loop diagrams of the model, and the mass $M$ is taken at the infinite limit at the end of the calculation. It plays the role of a cutoff.  Note however, that, if the limit  $M\to \infty$  is not taken in the above formula, then the introduction of such massive scalar particle   gives a contribution with a minus sign to the propagator. An standard scalar particle instead should give a plus sign contribution. Therefore, the Pauli-Villars procedure introduces a particle with wrong sign in the kinetic term, for this reason this particle is considered as a ghost. As the mass is taken $M\to\infty$ this ghost decouples, which softens the problem. If instead this limit is not taken,  the desired behavior $k^{-4}$ for the propagator is obtained at the cost of the introduction of a massive state with mass $M$ with wrong kinetic term. 

Note that theory of two scalar fields, such as the one resulting from the Pauli-Villars method before taking the $M\to \infty$ limit may  be interpreted, in a very loose way,  as the theory of one scalar field with fourth order derivatives, obtained by the standard mathematical trick of doubling degrees of freedom in order to lower the order of the resulting equations of motion. This is the analog of the standard procedure for going to the Euler-Lagrange equations to the Hamilton ones in Classical Mechanics.

These considerations lead to consider generalizations of GR with quartic order equations of motion. Examples are known in the classical literature. For instance, the equations of motion of the quadratic action \cite{stelle1}-\cite{stelle2}
\begin{equation}\label{geld}
S_g=\frac{1}{16\pi G_N}\int \bigg[R+\alpha R^2+\beta R_{\mu\nu}R^{\mu\nu}\bigg]\sqrt{-g}d^4x,
\end{equation}
or even the ones for generic $F(R)$ theories, are of fourth order, when varied by employing the metric formalism. In addition, the scales $\alpha G^{-1}_N$ and $\beta G_N^{-1}$ are dimensionless, which implies that the $n$-graviton vertices have to be quartic in momentum, as desired for the right power counting introduced above. This motivates the focus on these type of theories, which is the motivation of the following sections. The discussion of the previous paragraph however, anticipates that there may be a ghost in the spectrum. There is in fact, as shown in detail in \cite{stelle1}.

Before studying these topics, it should be mentioned that the arguments given above are plausible, however, the renormalization procedure has to take into account that the gauge fixing introduce graviton ghost couplings. Therefore, a more detailed analysis has to be done before to conclude renormalizability. This analysis was presented in \cite{stelle2}, where renormalization was established. The main purpose of the next section is to review these arguments but in addition, to give a further interpretation of the results of \cite{stelle2}.  

For specific details about gravity renormalization techniques, especially with loops present, an incomplete list is \cite{barth}-\cite{feynman} and references therein.
\subsection{The propagator in certain higher order gravity theories}

Before to studying the BRST symmetry and the Slavnov-Taylor equations for the Stelle model, it is convenient to give an expanded version of the features  about quartic gravity theories given in the previous subsection. Consider a generic model
\begin{equation}\label{tat}
S_g=\frac{1}{16\pi G_N}\int \bigg[R+R F_1(\square) R+R_{\mu\nu}F_2(\square)R^{\mu\nu}+R_{\mu\nu\alpha\beta}F_3(\square)R^{\mu\nu\alpha\beta}\bigg]\sqrt{-g}d^4x +S_{gf}.
\end{equation}
Here $S_{gf}$ is a gauge fixing term,  $\square$ denotes the wave D'Alambert operator on the background, and $F_i(\square)$ are functions of this operator. This type of models include the renormalizable model of Stelle \cite{stelle1}-\cite{stelle2} or the super-renormalizable models introduced in \cite{modesto1}-\cite{modesto4} as particular cases. 

The GR limit of the above theory is $F_i(\square)\to 0$. The Stelle limit is $F_1(\square)\to \alpha$, $F_2(\square)\to \beta$ and $F_3(\square)\to 0$.
The propagator $D_{\mu\nu,\alpha\beta}$ corresponding to a perturbation  $g_{\mu\nu}=\eta_{\mu\nu}+h_{\mu\nu}$ around flat space can be calculated by retaining the quadratic terms in $h_{\mu\nu}$ in the gravitational action. After some standard calculation, it follows that the quadratic terms for the model given above are collected as \cite{biswas}
$$
S^2_g=\frac{1}{32\pi G_N}\int \bigg[\frac{1}{2}h^{\alpha\beta}\eta_{\alpha\mu}\eta_{\beta\nu} \square a(\square) h^{\mu\nu}+h^{\alpha\beta}\eta_{\beta\mu}  b(\square)\partial_\alpha\partial_\nu h^{\mu\nu}+h   c(\square)\partial_\mu\partial_\nu  h^{\mu\nu}
$$
\begin{equation}\label{cuadrtizada}
+\frac{1}{2}h   \square d(\square)  h+h^{\alpha\beta}   \frac{f(\square)}{\square}\partial_\alpha\partial_\beta\partial_\mu\partial_\nu  h^{\mu\nu}\bigg] d^4x+S_{gf}.
\end{equation}
In the last expressions, the quantities
$$
a(\square)=1-\frac{1}{2} F_2(\square)\square-2F_3(\square)\square+a_{gf},\qquad 
b(\square)=-1+\frac{1}{2} F_2(\square)\square+2F_3(\square)\square+b_{gf},
$$
$$
c(\square)=1+2 F_1(\square)\square+\frac{1}{2}F_2(\square)\square+c_{gf},\qquad
d(\square)=-1-2 F_1(\square)\square-\frac{1}{2}F_2(\square)\square+d_{gf},
$$
$$
f(\square)=-2 F_1(\square)\square-F_2(\square)\square-2F_3(\square)\square+f_{gf},
$$
were introduced, where subscript "gf" indicates gauge fixing terms. A usual choice is the De Donder gauge, which is defined by
$$
\partial_\mu h^\mu_\nu=\frac{1}{2}\partial_\nu h^\mu_\mu.
$$
The  GR limit  corresponds to $F_i(\square)\to 0$. This limit, for the De Donder gauge, after the redefinition $h_{\mu\nu}\to \sqrt{16\pi G_N}h_{\mu\nu}$ should lead to the well known GR linearized gauge fixed action
$$
S_{GR}=\int \bigg[\frac{1}{4}h^{\alpha\beta}\eta_{\alpha\mu}\eta_{\beta\nu} \square h^{\mu\nu}
-\frac{1}{8}h   \square  h\bigg] d^4x.
$$
This leads to the conclusion that the gauge fixed coefficients are in this case
$$
a_d(\square)=1-\frac{1}{2} F_2(\square)\square-2F_3(\square)\square,\qquad 
b_d(\square)=\frac{1}{2} F_2(\square)\square+2F_3(\square)\square,
$$
$$
c_d(\square)=2 F_1(\square)\square+\frac{1}{2}F_2(\square)\square,\qquad
d_d(\square)=-\frac{1}{2}-2 F_1(\square)\square-\frac{1}{2}F_2(\square)\square,
$$
\begin{equation}\label{donde}
f_d(\square)=-2 F_1(\square)\square-F_2(\square)\square-2F_3(\square)\square.
\end{equation}
In the Stelle limit all these quantities in the momentum space are quadratic in momenta, since the $F_i(\square)$ are simple constants multiplied by $\square$.

The calculation of the propagator of the model requires to express the action \eqref{cuadrtizada} as in \eqref{tat}. This task may be achieved with the help of the Barnes -Rivers operators \cite{barnes}-\cite{barnes2}, which  are given in terms of the elementary tensors,
\begin{equation}\label{ripol2}
\omega_{\mu\nu}=\frac{k_\mu k_\nu}{k^2}, \qquad \theta_{\mu\nu}=\eta_{\mu\nu}-\frac{k_\mu k_\nu}{k^2},
\end{equation}
in the following way
$$
P^2_{\alpha \beta,\mu\nu}=\frac{1}{2}(\theta_{\beta \mu}\theta_{\alpha\nu}+\theta_{\beta\nu}\theta_{\alpha\mu})-\frac{1}{3}\theta_{\beta\alpha}\theta_{\mu\nu},
$$
$$
P^1_{\alpha \beta,\mu\nu}=\frac{1}{2}(\theta_{\beta\mu}\omega_{\alpha\nu}+\theta_{\beta\nu}\omega_{\alpha\mu}+\theta_{\alpha\mu}\omega_{\beta\nu}+\theta_{\alpha\nu}\omega_{\beta\mu}),
$$
$$
P^{0-s}_{\alpha \beta,\mu\nu}=\frac{1}{3}\theta_{\beta\alpha}\theta_{\mu\nu},\qquad
P^{0-w}_{\alpha \beta,\mu\nu}=\omega_{\beta\alpha}\omega_{\mu\nu},
$$
\begin{equation}\label{ripol}
P^{0-sw}_{\alpha \beta,\mu\nu}=\frac{1}{\sqrt{3}}\theta_{\beta\alpha}\omega_{\mu\nu},\qquad
P^{0-ws}_{\alpha \beta,\mu\nu}=\frac{1}{\sqrt{3}}\omega_{\beta\alpha}\theta_{\mu\nu}.
\end{equation}
In the coordinate space
$$
\omega_{\mu\nu}=\frac{\partial_\mu \partial_\nu}{\square}, \qquad \theta_{\mu\nu}=\eta_{\mu\nu}-\frac{\partial_\mu \partial_\nu}{\square}.
$$
Given an expression of the form
$$
M=a_2 P^2+a_1 P^1+a_s P^{0-s}+a_w P^{0-w}+a_{sw}\sqrt{3}(P^{0-sw}+P^{0-ws}),
$$
its inverse is  \cite{julves}
\begin{equation}\label{inverso}
M^{-1}=\frac{1}{a_2} P^2+\frac{1}{a_1} P^1+\frac{1}{a_s a_w-3 a_{sw}^2}\bigg[a_w P^{0-s}+a_s P^{0-w}-a_{sw}\sqrt{3}(P^{0-sw}+P^{0-ws})\bigg].
\end{equation}
The last formula is useful for the calculation of the  propagator of the model. The application of these formulas is as follows. First. note that by going to the momentum space, the first term of the action \eqref{cuadrtizada} becomes
$$
h^{\alpha\beta}\eta_{\alpha\mu}\eta_{\beta\nu} \square a(\square) h^{\mu\nu}\to -k^2a(-k^2)h^{\alpha\beta}\eta_{\alpha\mu}\eta_{\beta\nu}  h^{\mu\nu}.
$$
By employing the definition \eqref{ripol2}, it is evident that $\eta_{\mu\nu}=\theta_{\mu\nu}+\omega_{\mu\nu}$, and with this simple identity the last term in momentum space may be worked out as
$$
-k^2a(-k^2)h^{\alpha\beta}\eta_{\alpha\mu}\eta_{\beta\nu}  h^{\mu\nu}=-k^2a(-k^2)h^{\alpha\beta}(\theta_{\alpha\mu}\theta_{\beta\nu}+\theta_{\alpha\mu}\omega_{\beta\nu}+\omega_{\alpha\mu}\theta_{\beta\nu}+\omega_{\alpha\mu}\omega_{\beta\nu})  h^{\mu\nu}
$$
$$
=-k^2a(-k^2)h^{\alpha\beta}(P^{2}_{\alpha\beta, \mu\nu}+P^{0}_{\alpha\beta, \mu\nu}+P^{1}_{\alpha\beta, \mu\nu}+P^{0-w}_{\alpha\beta, \mu\nu})  h^{\mu\nu}.
$$
In the last identity, the definition of the Barnes Rivers operators \eqref{ripol} was employed. The analogous procedure may be applied for the remaining terms of the action. The result is
$$
S^2_{gf}=\int h^{\alpha\beta} D^{-1}_{\alpha\beta,\mu\nu} h^{\mu\nu} d^4x,
$$
where the quantity $D^{-1}_{\alpha\beta,\mu\nu}$ in momentum space reads in terms of the gauge fixed coefficients \eqref{donde} as follows
$$
D^{-1}_{\alpha\beta, \mu\nu}=-k^2\bigg[\frac{a}{2} P^2+\frac{a+b}{2} P^1+\frac{a+3d}{2} P^{0-s}
$$
$$
+\frac{a+b+2c+d+2f}{2} P^{0-w}+\frac{\sqrt{3}(c+d)}{2}(P^{0-sw}+P^{0-ws})\bigg]_{\alpha\beta,\mu\nu}.
$$
 The propagator in De Donder gauge is the inverse of the last expression, which can be found easily by use of \eqref{inverso}. The result is
$$
D_{\alpha\beta, \mu\nu}=-\frac{1}{k^2}\bigg[\frac{2}{a} P^2+\frac{2}{a+b} P^1+\frac{a+b+2c+d+2f}{2D} P^{0-s}
$$
$$
+\frac{a+3d}{2D} P^{0-w}-\frac{\sqrt{3}(c+d)}{2D}(P^{0-sw}+P^{0-ws})\bigg]_{\alpha\beta,\mu\nu}.
$$
Here, the following quantity $$
4D=(a+3d)(a+b+2c+d+2f)-(c+d)^2,$$
has been introduced.

The expressions given above can be simplified further by using \eqref{donde}, which implies that
$a+b=1$,  $c+d=-\frac{1}{2}$  and $b+c+f=0$. These relations allows to express the quantities of interest in terms of two of the functions, which may be  chosen as $a$ and $d$, resulting in
$$
Q_{\alpha\beta, \mu\nu}=-k^2\bigg[\frac{a}{2} P^2+\frac{1}{2} P^1+\frac{a+3d}{2} P^{0-s}
+\frac{2a+d-1}{2} P^{0-w}-\frac{\sqrt{3}}{4}(P^{0-sw}+P^{0-ws})\bigg]_{\alpha\beta,\mu\nu},
$$
\begin{equation}\label{propagador}
D^{\alpha\beta, \mu\nu}=-\frac{1}{k^2}\bigg[\frac{2}{a} P^2+ 2P^1+\frac{2a+d-1}{2D} P^{0-s}
+\frac{a+3d}{2D} P^{0-w}+\frac{\sqrt{3}}{4D}(P^{0-sw}+P^{0-ws})\bigg]_{\alpha\beta,\mu\nu},
\end{equation}
with
$$D=\frac{1}{4}(a+3d)(2a+d-1)-\frac{3}{16}.$$
The quantity \eqref{propagador} will be fundamental for studying quantum gravity processes.  As a consistency test, it may be noticed that the GR limit, which follows by imposing $F_i(\square)=0$, implies by \eqref{donde} that $a\to 1$ and $2d\to -1$. In this limit, the last expressions reduce to
\begin{equation}\label{relativiza}
Q_{\alpha\beta,\mu\nu}=-\frac{k^2}{2}\eta_{\alpha\mu}\eta_{\beta\nu}+\frac{k^2}{4}\eta_{\alpha\beta}\eta_{\mu\nu},\qquad 
D_{\alpha\beta,\mu\nu}=-\frac{1}{2k^2}\bigg[\eta_{\alpha\mu}\eta_{\beta\nu}+\eta_{\alpha\nu}\eta_{\beta\mu}-\eta_{\alpha\beta}\eta_{\mu\nu}\bigg],
\end{equation}
which are the known quantities  corresponding to GR in the De Donder gauge. This propagator goes of course
as $k^{-2}$.

In addition, note that in the Stelle limit described below \eqref{tat} the quantity $a(k)$ is given by a quadratic expression of the form $a(k)=u-v k^2$, as remarked below \eqref{donde}. The same follows for $d(k)$. This implies that there are terms in the propagator $D^{\alpha\beta, \mu\nu}$ that go like $k^{-4}$, which is a desired feature by power counting. The attentive reader however, may notice that the presence of terms that go like $k^{-4}$ is not enough for a good power counting. The calculation given above depends on the choice of gauge, and depending on this choice, terms that fall slower than $k^{-4}$ may appear. This problem will be discussed below, although it can be already mentioned that there are gauge fixing terms that do not spoil this desired behavior.

The calculation of the interaction vertices of the above theory is more involved \cite{prinz},  however they are of fourth order in the momentum of the particles pointing to that vertex. The methods developed in \cite{prinz} are suitable for their explicit calculation, and we refer the reader to this reference for further information.

\subsection{Gravitational BRST symmetries}
In general gravity theories, the transformation \eqref{infdif} plays the role of the gauge transformation of the model, with the gauge field identified by $h^{\mu\nu}$. In ordinary non abelian gauge theories, the gauge transformation on the gauge field is given by $$g\delta_\xi A_\mu^a=D_\mu^{ab} \xi^b,$$with $D_\mu^{ab}$ the covariant derivative for the gauge field.  Consider instead  a generic gravity action $L_g$ with a gauge fixing $\partial_\nu h^{\nu\mu}=0$.  By introducing $F^\tau_{\mu\nu}=\delta_\mu^\tau \partial_\nu$, the gauge fixed action  becomes
$$
S_{gf}=\int d^4x[L_s+\overline{\eta}_\tau F_{\mu\nu}^\tau D^{\mu\nu}_{\alpha}\eta^\alpha].
$$
As for Yang-Mills theories, a gauge fixing term can be added by an appropriate weight functional in the Feynman integral such as in \eqref{feynman2}. This procedure, adapted to the present case, leads to the gauge fixed action 
$$
S_{gf}=\int d^4x[L_s+\overline{\eta}_\tau F_{\mu\nu}^\tau D^{\mu\nu}_{\alpha}\eta^\alpha-\frac{\kappa^2}{2\lambda}F_\tau F^\tau].
$$
Here $F^\tau=F^\tau_{\mu\nu} h^{\mu\nu}$.   However, as pointed out in the previous subsection, the introduction of such term may spoil the  desired behavior $k^{-4}$ leading to \eqref{nocrece}. This feature does not necessarily spoils renormalizability, but makes the topic much harder. For this and other  reasons, the original reference \cite{stelle2} employ another weight functional inside the path integral leading to
\begin{equation}\label{deaction}
S_{gf}=\int d^4x[L_s-\frac{\kappa^2}{2\lambda}F_\tau\square F^\tau+\overline{\eta}_\tau F_{\mu\nu}^\tau D^{\mu\nu}_{\alpha}\eta^\alpha].
\end{equation}
For this choice, the resulting Green functions fall no slower than $k^{-4}$, as the gauge fixing term is of fourth order. 

As for Yang-Mills fields, the gauge fixed action $S_{gf}$ is not invariant under the infinitesimal diffeomorphisms  \eqref{infdif}.  In fact, it should not be so, as it reflects a particular choice of gauge. The task is to find a transformation, if it exists, leaving the gauge fixed action unchanged, perhaps up to a boundary term. This is the analog to the BRST symmetry described in \eqref{sobran}-\eqref{s}. It is natural to try to generalize the procedure described in those formulas adapted to the present case. Instead of going through all the calculations  as for the Yang-Mills case, one may employ the gained knowledge in those sections. A try may be  to find a BRST transformation that increases the ghost number by one, supposing that these numbers are the same as for the Yang-Mills counterpart. In fact they are, as the assignation of the same ghost numbers than in Yang-Mills theories leads to a zero ghost number to the action $S_{gf}$, which is a desired feature. The ansatz for BRST transformation is then
\begin{equation}\label{brstgrav}
\delta h^{\mu\nu}=\kappa D_\alpha^{\mu\nu}\eta^\alpha\epsilon,\qquad \delta\eta^\alpha=-\kappa^2 \partial_\beta \eta^\alpha \eta^\beta\epsilon,\qquad \delta\overline{\eta}_\tau=-\frac{\kappa^3}{\lambda} \square^2 F_\tau\epsilon.
\end{equation}
This trial clearly increase the ghost number by one. The expression of $\delta h^{\mu\nu}$ is simple the action of  diffeomorphisms, with $\eta^\alpha\epsilon$ as a parameter. Here $\epsilon $ is taken as anti-commuting. The reason for the derivative in $\delta\eta^\alpha$, which is not present in the Yang-Mills theories in \eqref{brstgrassmann}, is the following. In Yang-Mills theories, $\eta^a$ is an scalar and the index $a$ is a gauge group index. In the present case, $\eta^\alpha$ is a vector.  If one increases the ghost number by one, then a quadratic expression is needed, and the index $\beta$ has to be saturated with the derivative $\partial_\beta$.  Besides, it is clear due to the anti-commuting nature of $\eta^\alpha$ and $\epsilon$ that,  by defining $s$ as the part of \eqref{brstgrav} independent of $\epsilon$, the following formula is true
$$
\delta s\eta^\alpha=k^4\partial_\beta(\partial_\gamma \eta^\alpha \eta^\gamma \epsilon) \eta^\beta+\kappa^4\partial_\beta \eta^\alpha \partial_\gamma \eta^\beta \eta^\gamma \epsilon
$$
$$
=-\kappa^2\partial_\beta \partial_\gamma(\eta^\alpha)\eta^\gamma \eta^\beta\epsilon-\kappa^2\partial_\gamma \eta^\alpha \partial_\beta \eta^\gamma \eta^\beta \epsilon+\kappa^2\partial_\beta \eta^\alpha \partial_\gamma \eta^\beta \eta^\gamma\epsilon=0.
$$
The last identity is due to the fact that the first term is a product of symmetric by anti-symmetric, as $\eta^\gamma \eta^\beta=-\eta^\beta\eta^\gamma$, while the second and the third clearly cancel each other. Thus the BRST action \eqref{brstgrav}
is nilpotent when acting on $\eta^\alpha$, that is, $s^2 \eta^\alpha=0$.

Besides, the last term $\delta\overline{\eta}_\tau$ is similar to the ones in Yang-Mills theories, described below \eqref{sobran}.  In fact, the  action of \eqref{brstgrav} on the ghost part of the lagragian in   \eqref{deaction}  can be expressed as
$$
\delta(-\frac{\kappa^2}{2\lambda}F_\tau\square F^\tau+\overline{\eta}_\tau F_{\mu\nu}^\tau D^{\mu\nu}_{\alpha}\eta^\alpha)=-\frac{\kappa^3}{2\lambda}F^\tau_{\mu\nu} D^{\mu\nu}_\alpha \eta^\alpha\epsilon\square F^\tau-\frac{\kappa^3}{2\lambda}F^\tau\square F^\tau_{\mu\nu} D^{\mu\nu}_\alpha \eta^\alpha\epsilon
$$
\begin{equation}\label{deaction2}
-\frac{\kappa^3}{\lambda} (\square^2F_\tau\epsilon) F_{\mu\nu}^\tau D^{\mu\nu}_{\alpha}\eta^\alpha+\overline{\eta}_\tau F_{\mu\nu}^\tau \delta(D^{\mu\nu}_{\alpha}\eta^\alpha).
\end{equation}
The three first terms of the right cancel up to a total derivative, due to the anti commuting nature of $\eta^\alpha$ and $\epsilon$.
Therefore BRST invariance \eqref{brstgrav} will be ensured if the last term vanish, which can be realized if and only if
$$
\delta(D^{\mu\nu}_{\alpha}\eta^\alpha)=0.
$$
In other words, the action of $s$ on the metric $h^{\mu\nu}$ has to be nilpotent as well. This is pretty similar to the situation for Yang-Mills theories, described in \eqref{musvanish}. In proving that formula, the fundamental property  was the Jacobi identity for the gauge algebra constants $c_{abc}$.
These constants are given by the commutator of gauge group elements. This suggest that, in the present situation, the commutator of $D_\alpha^{\mu\nu}$ may play this important role. This can be checked explicitly. The left hand of the last condition is equivalent to
$$
\delta(D^{\mu\nu}_{\alpha}\eta^\alpha)=-\kappa^2D^{\mu\nu}_{\alpha}\partial_\gamma\eta^\alpha\eta^\gamma\epsilon+\kappa\frac{\delta D_\gamma^{\mu\nu}}{\delta h^{\alpha\beta}}D_\epsilon^{\alpha\beta} \eta^\epsilon \eta^\gamma\epsilon
$$
\begin{equation}\label{mysweetlord}
=-\kappa^2 D^{\mu\nu}_{\alpha}\partial_\gamma\eta^\alpha\eta^\gamma\epsilon+\frac{\kappa}{2}\bigg(\frac{\delta  D_\epsilon^{\mu\nu}}{\delta h^{\alpha\beta}}D_\gamma^{\alpha\beta}-\frac{\delta D_\gamma^{\mu\nu}}{\delta h^{\alpha\beta}}D_\epsilon^{\alpha\beta}\bigg) \eta^\epsilon \eta^\gamma\epsilon,
\end{equation}
where in the last step  the property $\eta^\alpha\eta^\gamma=-\eta^\gamma \eta^\alpha$ was employed. On the other hand, the commutator of two diffeomorphisms with parameters $\xi^\alpha$ and $\eta^\alpha$, not necessarily Grassman ones,  can be found by taking  \eqref{infdif}  and by calculating
$$
(\delta_\eta \delta_\xi-\delta_\xi \delta_\eta) h^{\mu\nu}=\bigg(\frac{\delta D_\epsilon^{\mu\nu}}{\delta h^{\alpha\beta}}D_\gamma^{\alpha\beta}-\frac{\delta D_\gamma^{\mu\nu}}{\delta h^{\alpha\beta}}D_\epsilon^{\alpha\beta}\bigg)\eta^{\epsilon}\xi^\gamma.
$$
This expression, by use of the explicit definition \eqref{infdif} and by taking $\xi^\alpha=\eta^\alpha$ with $\eta^\alpha$ a Grassman variable, can be shown after a direct, although slightly tedious calculation, to be equal to 
$$
\bigg(\frac{\delta D_\epsilon^{\mu\nu}}{\delta h^{\alpha\beta}}D_\gamma^{\alpha\beta}-\frac{\delta D_\gamma^{\mu\nu}}{\delta h^{\alpha\beta}}D_\epsilon^{\alpha\beta}\bigg) \eta^\alpha \eta^\gamma=2\kappa D^{\mu\nu}_{\alpha}\partial_\gamma\eta^\alpha\eta^\gamma.
$$
The last identity, together with \eqref{mysweetlord} implies that $\delta(D^{\mu\nu}_{\alpha}\eta^\alpha)=\delta s h^{\mu\nu}=0$, in other words, the BRST action is nilpotent when acting on the gravitational perturbation $h^{\mu\nu}$. 

In brief, the above discussion shows that the gauge fixed action \eqref{deaction} is invariant under the BRST action \eqref{brstgrav}, which is nilpotent acting on the metric field $h^{\mu\nu}$ and on the ghost $\eta^\alpha$.

\subsection{The effective gravitational action}
In analogy with the procedure outlined in \eqref{conexa}, the  effective action   $G(T, \xi, \overline{\xi}, K, L)$  for gravitational theories, generating the connected correlation functions, is defined by the following relation
$$
e^{i G( \xi, \overline{\xi}, K, L, T)}=\int Dh^{\mu\nu}D\eta^a D\overline{\eta}^b \exp\{i\int d^4x[\overline{\xi}_\alpha\eta^\alpha+\xi_\alpha\overline{\eta}^\alpha+\kappa J_{\mu\nu}h^{\mu\nu}]\}
$$
\begin{equation}\label{eliana}
\exp\{i\int d^4x[L_s-\frac{\kappa^2}{2\lambda}F_\tau\square F^\tau+\overline{\eta}_\tau F_{\mu\nu}^\tau D^{\mu\nu}_{\alpha}\eta^\alpha+\kappa K_{\mu\nu}D^{\mu\nu}_\alpha \eta^\alpha +\kappa^2 L_\alpha \partial_\beta \eta^\alpha \eta^\beta]\}.
\end{equation}
Here the current  $J_{\mu\nu} h^{\mu\nu}$ is the analogous to the gauge field current coupling $J_\mu A^\mu$ for Yang-Mills theories, and $J_{\mu\nu}$ may be identified as the energy momentum tensor $J_{\mu\nu}=T_{\mu\nu}$ for an external source. As in those theories, the sources $K_{\mu\nu}$ and $L_\alpha$ are coupled to the nilpotent part of the BRST invariance. From this expression it follows that the gauge fixed action for the gravitational perturbation can be expressed as follows
$$
S_{gf}=\int d^4x[L_s-\frac{\kappa^2}{2\lambda}F_\tau\square F^\tau+\overline{\eta}_\tau F_{\mu\nu}^\tau D^{\mu\nu}_{\alpha}\eta^\alpha+\kappa K_{\mu\nu}D^{\mu\nu}_\alpha \eta^\alpha +\kappa^2 L_\alpha \partial_\beta \eta^\alpha \eta^\beta].
$$
The mass dimensions of the fields are directly seen from this action 
\begin{equation}\label{masaso}
[h^{\mu\nu}]=0,\qquad [\eta^a]=[\overline{\eta}^a]=1, \qquad [K_{\mu\nu}]=[L_\alpha]=3.
\end{equation}
Here, as before, the values for the ghost and anti-ghost are non uniquely defined. The choice above is however valid, and is the one that is usually employed.

Another important issue is how the quantities above transform under Lorentz transformation. The fields $h^{\mu\nu}$ is a second order contravariant tensor.  while the ghost $\eta_{\alpha}$ and $\overline{\eta}_{\alpha}$ are covariant vectors. This is different than in gauge theories, where they were scalars. The lagrangian should be a Lorenz scalar, which leads to the conclusion $K_{\mu\nu}$ is a covariant tensor  and $L_a$ is a covariant vector. 

The  ghost symmetry of the model given by
\begin{equation}\label{ghostsys}
\eta^\beta\to e^{i\alpha}\eta^\beta,\qquad  \overline{\eta}^\beta\to e^{-i\alpha}\overline{\eta}^\beta, \qquad \xi_\beta\to e^{-i\alpha}\xi_\beta,\qquad  \overline{\xi}_\beta\to e^{i\alpha} \overline{\xi}_\beta,
\end{equation}
$$
K_{\mu\nu}\to e^{-i\alpha}K_{\mu\nu},\qquad L_\beta\to e^{-2i\alpha} L_\beta,
$$
with corresponding ghost number
\begin{equation}\label{Qs}
Q(h^{\mu\nu})=Q(J_{\mu\nu})=0,\qquad Q(L_\alpha)=-2
\end{equation}
$$
Q(\eta^\alpha)=-Q(\overline{\eta}^\alpha)=-Q(\xi_\alpha)=Q(\overline{\xi}_\alpha)=- Q(K_{\mu\nu})=1.
$$
This is the analogous of \eqref{Q}   for Yang-Mills theories adapted to the present case. The quantum action, analogous to \eqref{quantumaction}, for the gravitational case is simply
$$
\Gamma( <h>, <\eta>,<\overline{\eta}>, K, L)=-i G(J, \xi, \overline{\xi}, K, L)
$$
\begin{equation}\label{quantumactions}
-\int d^4 x[J_{\mu\nu}  <h^{\mu\nu}>+\overline{\xi}_a<\eta^a>+\xi_a<\overline{\eta}^a>].
\end{equation}
 The BRST invariance of the action $S_{gf}$ leads to  Slavnov-Taylor identities such as \eqref{slavnov}, which for the present case read
\begin{equation}\label{slavnovs}
\int d^4x\bigg[\frac{\delta \Gamma}{\delta  h^{\mu\nu} }\frac{\delta \Gamma}{\delta K_{\mu\nu}}+\frac{\delta  \Gamma}{\delta \eta^\alpha}\frac{\delta \Gamma}{\delta L_\alpha}\bigg]=0.
\end{equation}
The anti-ghost translation $\overline{\eta}^\alpha\to \overline{\eta}^\alpha+\delta\overline{\eta}^\alpha$  is reflected in the ghost equation of motion
\begin{equation}\label{ghostmotions}
F_{\mu\nu}^\tau \frac{\delta \Gamma}{\delta K_{\mu\nu}}+\frac{\delta \Gamma}{\delta \overline{\eta}_\tau}=0,
\end{equation} 
which is the analogous to the ghost equation \eqref{ghostmotion} adapted to gravitation. In the last two formulas, the redefinition $\Gamma\to\Gamma+\int d^4x\frac{\kappa^2}{2\lambda}F_\tau\square F^\tau$ has been made. The last formula leads to the  functional dependence
\begin{equation}\label{leider}
\Gamma=\Gamma(\kappa K_{\alpha\beta}-\overline{\eta}_\tau \overline{F}_{\alpha\beta}^\tau, \eta_\alpha, L_\alpha, h^{\mu\nu}), 
\end{equation}
where $\overline{F}_{\alpha\beta}^\tau$ indicates the operator $F_{\alpha\beta}^\tau$ acting on the  left.  

Until now, the reader may notice that the discussion is pretty similar to Yang-Mills theories.  The task is now to study the solutions of the above derived gravitational Slavnov-Taylor identities.

\subsection{Consequences of gravitational Slavnov-Taylor identiies}
Under the assumption, analogous to the one described for gauge theories in previous section, that the effective action admits an expansion of the form
$$
\Gamma=\Gamma_0+\hbar\Gamma_1+\hbar^2\Gamma_2+...
$$
the Slavnov-Taylor identity \eqref{slavnovs}, truncated to order $n$, is given by
$$
\sum_{p+q=n} \int d^4x\bigg[\frac{\delta \Gamma_p}{\delta  h^{\mu\nu} }\frac{\delta \Gamma_q}{\delta K_{\mu\nu}}+\frac{\delta  \Gamma_p}{\delta \eta^\alpha}\frac{\delta \Gamma_q}{\delta L_\alpha}\bigg]=0.
$$
This in fact is true  for $p+q\leq n$. The zero action is $\Gamma_0(S_{(0)gf})=S_{(0)gf}$ where
\begin{equation}\label{austria}
S_{(0)gf}=\int d^4x[L_s+\overline{\eta}_{0\tau} F_{\mu\nu}^\tau D^{\mu\nu}_{\alpha}\eta_0^\alpha+\kappa K_{0\mu\nu}D^{\mu\nu}_\alpha \eta_0^\alpha +\kappa^2 L_{0\alpha} \partial_\beta \eta_0^\alpha \eta_0^\beta].
\end{equation}
This is the gauge fixed action described above, the $(0)$ index is just indicating that this is the initial piece fo an iterative procedure, to be described now. This initial piece fulfills the Slavnov-Taylor identity
\begin{equation}\label{i0s}
\int d^4x\bigg[\frac{\delta S_{(0)gf}}{\delta  h^{\mu\nu} }\frac{\delta S_{(0)gf}}{\delta K_{\mu\nu}}+\frac{\delta  S_{(0)gf}}{\delta \eta^\alpha}\frac{\delta S_{(0)gf}}{\delta L_\alpha}\bigg]=0.
\end{equation}
The next order effective action is  $\Gamma=\Gamma_0+\hbar \Gamma_1=S_{(0)gf}+\hbar \Gamma_1$, where $\Gamma_1$ is to be constrained by the Slavnov-Taylor equation
$$
\int d^4x\bigg[\frac{\delta S_{(0)gf}}{\delta  h_0^{\mu\nu}}\frac{\delta \Gamma_1}{\delta K_{0\mu\nu}}+\frac{\delta  S_{(0)gf}}{\delta \eta_0^\alpha}\frac{\delta \Gamma_1}{\delta L_{0\alpha}}+\frac{\delta \Gamma_1}{\delta  h_0^{\mu\nu} }\frac{\delta S_{(0)gf}}{\delta K_{0\mu\nu}}+\frac{\delta  \Gamma_1}{\delta \eta_0^\alpha}\frac{\delta S_{(0)gf}}{\delta L_{0\alpha}}\bigg]=0.
$$
This leads to an equation of the form
$$
\sigma_0\Gamma_1(S_{(0)gf})=\partial_\mu D^\mu,
$$
where the term $D^\mu$ inside the divergence has to have suitable vanishing conditions at the boundary of the space time, and  where the nilpotent operator 
\begin{equation}\label{putono}
\sigma_0=\frac{\delta S_{(0)gf}}{\delta  h_0^{\mu\nu}}\frac{\delta }{\delta K_{0\mu\nu}}+\frac{\delta  S_{(0)gf}}{\delta \eta_0^\mu}\frac{\delta }{\delta L_{0\mu}}+\frac{\delta S_{(0)gf}}{\delta  K_{0\mu\nu} }\frac{\delta }{\delta h_0^{\mu\nu}}+\frac{\delta  S_{(0)gf}}{\delta L_{0\mu}}\frac{\delta }{\delta \eta_0^\mu},
\end{equation}
has been introduced. The resulting equation becomes
\begin{equation}\label{sumoso}
\frac{\delta S_{(0)gf}}{\delta  h_0^{\mu\nu}}\frac{\delta \Gamma_1 }{\delta K_{0\mu\nu}}+\frac{\delta  S_{(0)gf}}{\delta \eta_0^\mu}\frac{\delta \Gamma_1 }{\delta L_{0\mu}}+\frac{\delta S_{(0)gf}}{\delta  K_{0\mu\nu} }\frac{\delta \Gamma_1 }{\delta h^{0\mu\nu}}+\frac{\delta  S_{(0)gf}}{\delta L_{0\mu}}\frac{\delta \Gamma_1}{\delta \eta_0^\mu}=\partial_\mu D^\mu.
\end{equation}
In the section of Yang-Mills theories, two methods for solving equations of this type were presented. One is solving this equation by brute force, the other employs the fact that $\sigma_0$ is nilpotent, as explained below \eqref{potonto}. In the following, this second method will be applied, as it leads to the correct result in much easier fashion.  Due to the nilpotency property of $\sigma_0$, it will be assumed the first order action $\Gamma_1$ solving \eqref{sumoso} can be written as $\Gamma_1=S+\int d^4x\sigma_0 X$ where $S$ is a diffeomorphism invariant functional of $h^{\mu\nu}$ and the functional $X$ has to have ghost number $-1$, and has to be invariant under Lorenz transformations.

Before applying that method, it should be mentioned that, at the best of the authors knowledge, there is no complete proof that such functional form works, although is plausible. There exist references such as \cite{henneaux}-\cite{anselmi3} which deeply study cohomological methods for solving this type of equations where several details may be consulted.  In any case, by taking  this functional form as granted, the invariance described above  reduce $X$ to the form
\begin{equation}\label{puton}
X=(\kappa K_{\mu\nu}-\overline{\eta}^\tau \overline{F}_{\mu\nu}^\tau)P^{\mu\nu}(h^{\alpha\beta})+L_\mu Q_{\nu}^{\mu}(h^{\alpha\beta}) \eta^\nu,
\end{equation}
where the quantities $P^{\mu\nu}(h^{\alpha\beta})$ and $Q_{\nu}^{\mu}(h^{\alpha\beta})$ may depend not only  on $h^{\alpha\beta}$ but on its derivatives as well, since for instance, $\kappa\partial_\gamma h^{\alpha\beta}$ is dimensionless and higher order derivatives may be rendered dimensionless by an appropriate power of $\kappa$. However, the analysis of Feynman diagrams with ghosts in \cite{stelle2}  shows that renormalizability is spoiled with the addition of these derivatives, as they increase the degree of divergence without limit. For these reasons the quantities $P^{\mu\nu}(h^{\alpha\beta})$ and $Q_{\nu}^{\mu}(h^{\alpha\beta})$ will be taken as functions of $h^{\alpha\beta}$ solely.

Given  the generic form for $X$ in \eqref{puton}, the next task is to find $\Gamma_1=S+\int d^4x\sigma_0 X$, which requires the calculation of $\sigma_0$.  This is not difficult. First, it is direct to find from \eqref{puton} that
$$
\frac{\delta X}{\delta K_{\mu\nu}}=\kappa P^{\mu\nu}(h^{\alpha\beta}),\qquad \frac{\delta X}{\delta \eta^\nu}=L_\mu Q_{\nu}^{\mu}(h^{\alpha\beta}) ,\qquad \frac{\delta X}{\delta L_\mu}=Q_{\nu}^{\mu}(h^{\alpha\beta}) \eta^\nu,
$$
\begin{equation}\label{prof}
\frac{\delta X}{\delta h^{\alpha\beta}}=(\kappa K_{\mu\nu}-\overline{\eta}^\tau \overline{F}_{\mu\nu}^\tau)\frac{\delta P^{\mu\nu}}{\delta h^{\alpha\beta}}+L_\mu \frac{\delta Q_{\nu}^{\mu}}{\delta h^{\alpha\beta}}\eta^\nu.
\end{equation}
In addition, a simple calculation employing \eqref{austria} leads to
$$
\frac{\delta S_{(0)gf}}{\delta K_{\mu\nu}}=\kappa D^{\mu\nu}_\alpha \eta_0^\alpha,\qquad \frac{\delta S_{(0)gf}}{\delta h^{\mu\nu}}=(\kappa K_{\alpha\beta}-\overline{\eta}^\tau \overline{F}_{\alpha\beta}^\tau)\frac{\delta D_\gamma^{\alpha\beta}}{\delta h_0^{\mu\nu}}\eta_0^\gamma+\frac{\delta S_0}{\delta h_0^{\mu\nu}},
 $$
 \begin{equation}\label{funcionales}
\frac{\delta S_{(0)gf}}{\delta L_0^\sigma}= \kappa^2 \partial_\alpha \eta_0^\sigma \eta_0^{\alpha} .\qquad \frac{\delta S_{(0)gf}}{\delta \eta_0^\sigma}=-(\kappa K_{\alpha\beta}-\overline{\eta}_0^\tau \overline{F}_{\alpha\beta}^\tau) D_\sigma^{\alpha\beta}+\kappa^2 L_{0\sigma} \eta_0^{\alpha}\overleftarrow{\partial}_\alpha-\kappa^2L_{0\alpha} \partial_{\sigma}\eta_0^\alpha.
 \end{equation}
Here the reversed arrow $\overleftarrow{}$ indicates the action of the operator on the left. With the help of the  last two formulas, the solution $\Gamma_1=S+\int d^4x\sigma_0 X$, can be found by use of  \eqref{puton} and \eqref{putono}, the result is
$$
\Gamma_1=\int d^4x\bigg[A(h_{\mu\nu})+\frac{\delta S_{0
}}{\delta h^{\mu\nu}}P^{\mu \nu}+(\kappa K_{\rho\sigma}-\overline{\eta}_\tau \overleftarrow{F}^\tau_{\rho\sigma})\bigg(\frac{\delta  D^{\rho\sigma}_\alpha}{\delta h^{\mu\nu}}\eta^\alpha P^{\mu\nu}-\frac{\partial P^{\rho\sigma}}{\delta h^{\mu\nu}} D^{\mu\nu}_\alpha\eta^\alpha - D^{\rho\sigma}_\alpha(Q^\alpha_\tau \eta^\tau)\bigg)
$$
\begin{equation}\label{diverge}
+\kappa L_\sigma \bigg(\kappa Q^\sigma_\tau\partial_\beta \eta^\tau-\kappa \partial_\beta(Q^\sigma_\tau \eta^\tau)-\kappa(\partial_\tau \eta^\sigma)Q^\tau_\beta -\frac{\delta Q^\sigma_\tau}{\delta h^{\mu\nu}}\eta^\tau D_\beta^{\mu\nu}\bigg) \eta^\beta\bigg].
\end{equation}
This is exactly the formula (6.17) of the original reference \cite{stelle2}. Here the following quantity 
$$
A(h_{\mu\nu})=\sqrt{-g}\bigg[\delta\kappa^{ 2}R+\delta\beta R^2+\delta\alpha R_{\mu\nu}R^{\mu\nu}\bigg],
$$
has been introduced and $S_{0}$ is the classical lagrangian of Stelle gravity. Note that  $A(h^{\mu\nu})$
 may be though as $$\int d^4x A(h^{\mu\nu})\sim S_0(\kappa+\delta \kappa, \alpha+\delta \alpha, \beta+\delta \beta, h^{\mu\nu})-S_0(\kappa, \alpha, \beta, h^{\mu\nu})
 $$
 $$
\sim \int d^4x\bigg[\delta\gamma\frac{\partial S_0}{\partial \gamma}+\delta\beta\frac{\partial S_0}{\partial \beta}+\delta\alpha\frac{\partial S_0}{\partial \alpha}\bigg],
 $$
if the quantities $\delta\kappa$, $\delta\alpha$ and $\delta\beta$ are considered as infinitesimal.  Here the constant $\gamma=\kappa^2$ was introduced.
The next task is to interpret all the terms in \eqref{diverge} in a more systematic manner.

\subsection{Physical interpretation of the counter terms for gravitational theories}

The effective action \eqref{diverge} collects the possible counter terms that can appear in quadratic gravity. It is a  solution of the Slavnov-Taylor identities. However, these identities alone do not insure that order by order, BRST invariance is preserved.  The purpose of the present section is to interpret the obtained counter terms in \eqref{diverge} and to elucidate their role in preserving BRST symmetry order by order.

For Yang-Mills theories, a tool for establishing multiplicative renormalization was the procedure outlined in \eqref{sanatorio}-\eqref{sanatorio2}, which consists in expressing the effective action in terms of functional derivatives of the gauge fixed functional $S_{(0)gf}$. Following the analogous procedure in gravity, with the help of the formulas \eqref{funcionales} leads to the following expression for the first order effective action \eqref{diverge} 
$$
\Gamma_1=\int d^4x\bigg[\delta\gamma\frac{\partial S_{(0)gf}}{\partial \gamma}+\delta\beta\frac{\partial S_{(0)gf}}{\partial \beta}+\delta\alpha\frac{\partial S_{(0)gf}}{\partial \alpha}-\frac{\delta S_{(0)gf} }{\delta h^{\mu\nu}}P^{\mu \nu}+\frac{\delta  S_{(0)gf}}{\delta \eta^\sigma}Q^\sigma_\tau \eta^\tau
$$
$$
-
(\kappa K_{\rho\sigma}-\overline{\eta}_\tau \overleftarrow{F}^\tau_{\rho\sigma})\frac{\partial P^{\rho\sigma}}{\delta h^{\mu\nu}} D^{\mu\nu}_\alpha\eta^\alpha +\kappa L_\sigma \bigg(\frac{\delta Q^\sigma_\tau}{\delta h^{\mu\nu}}\eta^\tau D_\beta^{\mu\nu}-Q^\sigma_\tau\partial_\beta( \eta^\tau)\bigg) \eta^\beta\bigg].
$$
Note that all the derivatives in this formula are expressed purely in terms of $S_{(0)gf}$, except for the last three terms. The last expression may be expressed as a functional tangent plane variation up to the mentioned last terms.  This suggests that
$$
\Gamma_1= S_{(0)gf}(\alpha+\delta \alpha, \beta+\delta\beta, \gamma+\delta \gamma, h^{\mu\nu}+P^{\mu\nu}, \eta^\alpha-Q^\alpha_\beta \eta^\beta, K_{\mu\nu}, \overline{\eta}_\alpha, L_\sigma)
$$
$$
- S_{(0)gf}(\alpha, \beta, \gamma, h^{\mu\nu}, \eta^\alpha, K_{\mu\nu}, \overline{\eta}_\alpha, L_\sigma)
$$
\begin{equation}\label{Qdf}
+\int d^4x\bigg[-
(\kappa K_{\rho\sigma}-\overline{\eta}_\tau \overleftarrow{F}^\tau_{\rho\sigma})\frac{\partial P^{\rho\sigma}}{\delta h^{\mu\nu}} D^{\mu\nu}_\alpha\eta^\alpha +\kappa L_\sigma \bigg(\frac{\delta Q^\sigma_\tau}{\delta h^{\mu\nu}}\eta^\tau D_\beta^{\mu\nu}-\kappa Q^\sigma_\tau\partial_\beta( \eta^\tau)\bigg) \eta^\beta\bigg],
\end{equation}
up to terms of order $\hbar^2$ or higher. 

The formula \eqref{Qdf} can be interpreted is the gravitational analogous of the one obtained in \eqref{sanatorio2}  and \eqref{reachingol} for the Yang-Mills case.  The main difference is that for the former case, the expression was fully given in terms of functional derivatives of the action $S_{(0)gf}$ while in \eqref{Qdf} there are three terms that are not given in that way. The first two terms suggest field redefinitions of the form $h^{\mu\nu}+P^{\mu\nu}$ and $ \eta^\alpha-Q^\alpha_\beta \eta^\beta$. 

The conclusions given above however, must be taken with care. The difference with Yang-Mills theories is  that in the later case, the field redefinitions involved multiplication constants $Z^n_i$ while the redefinition for gravity involves the  quantities $P^{\mu\nu}$ and $Q^\alpha_\beta$ which are arbitrary functions of $h^{\mu\nu}$. This makes things more subtle, as there may be implicit derivatives and a more careful analysis is needed.  

In any case, inspired by the above calculation, consider the possibility of  a field redefinition $h^{\mu\nu}\pm P^{\mu\nu}$ and $ \eta^\alpha\pm Q^\alpha_\beta \eta^\beta$. The quantities $Q_\alpha^\beta$ and $P^{\mu\nu}$ are considered of first order in $\hbar$.  The first subtle detail  is that,  given the action $S(h^{\mu\nu})$, the action  $S(h^{\mu\nu}+P^{\mu\nu}(h^{\alpha\beta}))$ with $P^{\mu\nu}(h^{\alpha\beta})$ an arbitrary function of $h^{\alpha\beta}$, does not describe the Stelle gravity anymore if  $h^{\alpha\beta}$ is still identified as the gravity perturbation. This is clear,  since if $P^{\alpha\beta}$ contains for instance a large power of $h^{\mu\nu}$ the resulting theory will not be a leading order of quadratic gravity. However, if one identifies instead $\widetilde{h}^{\mu\nu}=h^{\mu\nu}+P^{\mu\nu}(h^{\alpha\beta})$ as the true degree of freedom, the resulting theory will be the Stelle gravity.  On the other hand, \eqref{Qdf} suggests the redefinition  $\eta\to \widetilde{\eta}^a=\eta^\alpha\pm Q^\alpha_\beta(h^{\mu\nu})\eta^\beta$. Under a generic field redefinition $\widetilde{\eta}^\alpha=F^\alpha(\eta^\beta, h^{\gamma\delta})$, the BRST invariance \eqref{brstgrav} may be generalized to
\begin{equation}\label{brstgrav2}
\widetilde{s} \widetilde{h}^{\mu\nu}=\kappa \widetilde{D}_\alpha^{\mu\nu}\widetilde{\eta}^\alpha,\qquad \widetilde{s}\widetilde{\eta}^\alpha=-\kappa^2 \partial_\beta \widetilde{\eta}^\alpha \widetilde{\eta}^\beta.
\end{equation}
In light of these considerations,  the task is to analyze the terms in \eqref{diverge}. First note that under the redefinition $\widetilde{h}^{\mu\nu}=h^{\mu\nu}+P^{\mu\nu}(h^{\alpha\beta})$ and $\widetilde{\eta}^\alpha=\eta^\alpha-Q^\alpha_\beta \eta^\beta$ it is easy to deduce that
$$
\widetilde{\delta}(\widetilde{h}^{\mu\nu})-\delta(\widetilde{h}^{\mu\nu})=\widetilde{D}^{\mu\nu}_\alpha \widetilde{\eta}^\alpha \epsilon-D^{\mu\nu}_{\alpha} \eta^\alpha\epsilon-\frac{\partial P^{\mu\nu}}{\delta h^{\gamma\delta}} D^{\gamma\delta}_\alpha\eta^\alpha\epsilon
$$
\begin{equation}\label{coolo}
=\bigg(\frac{\delta D^{\mu\nu}_\alpha}{\delta h^{\rho\sigma}}\eta^\alpha P^{\rho\sigma}-D^{\mu\nu}_\alpha(Q^\alpha_\tau \eta^\tau)-\frac{\partial P^{\mu\nu}}{\delta h^{\rho\sigma}} D^{\rho\sigma}_\alpha\eta^\alpha\bigg)\epsilon.
\end{equation}
Here $\widetilde{\delta}$ is the BRST corresponding to \eqref{brstgrav2}  while $\delta$ corresponds \eqref{brstgrav}. In the last derivation a term proportional to a variation  of $P_{\mu\nu}$	times $Q^\alpha_\beta$ was neglected, as it is considered  a term of second order in $\hbar$. The reader will recognize, up to a factor $\epsilon$, that the last term in \eqref{coolo} corresponds to the factor that multiplies $(\kappa K_{\rho\sigma}-\overline{\eta}_\tau \overleftarrow{F}^\tau_{\rho\sigma})$ inside the solution of the Slavnov-Taylor identity \eqref{diverge}. 

In addition, the redefinition $\widetilde{\eta}^\alpha=\eta^\alpha-Q^\alpha_\beta \eta^\beta$ leads to the identities
$$
\widetilde{\delta}\widetilde{\eta}^\alpha-\delta \eta^\alpha=-\kappa^2[\partial_\beta \widetilde{\eta}^\alpha \widetilde{\eta}^\beta-\partial_\beta \eta^\alpha \eta^\beta]\epsilon=\kappa^2[\partial_\beta (Q^\alpha_\tau \eta^\tau)\eta^\beta+\partial_\beta( \eta^\alpha)Q^\beta_\tau \eta^\tau]\epsilon.
$$
 and 
$$
\delta\widetilde{\eta}^\alpha-\delta \eta^\alpha=-\delta(Q^\alpha_\beta \eta^\beta)=\kappa\bigg(\kappa Q^{\alpha}_\tau \partial_\beta \eta^\tau \eta^\beta-\frac{\delta Q^\alpha_\tau}{\delta  h^{\mu\nu}}\eta^\tau D^{\mu\nu}_\gamma \eta^\gamma\bigg)\epsilon.
$$
The subtraction of the last two expressions lead to
$$
\delta\widetilde{\eta}^\alpha-\widetilde{\delta}\widetilde{\eta}^\alpha
=[\kappa^2 Q^\alpha_\tau\partial_\beta \eta^\tau\eta^\beta-\kappa^2\partial_\beta(Q^\alpha_\tau \eta^\tau)\eta^\beta-\kappa^2(\partial_\tau \eta^\alpha)Q^\tau_\beta \eta^\beta-\kappa\frac{\delta Q^\alpha_\tau}{\delta h^{\mu\nu}}\eta^\tau D_\beta^{\mu\nu}\eta^\beta]\epsilon.
$$
Up to the $\epsilon$ factor, it is direct to see that, these are the sum of terms multiplying the factor $L_\sigma$
in \eqref{diverge}.  Therefore it is concluded that \eqref{diverge} can be expressed as
$$
\Gamma_1=\int d^4x\bigg[\delta\gamma\frac{\partial S_0}{\partial \gamma}+\delta\beta\frac{\partial S_0}{\partial \beta}+\delta\alpha\frac{\partial S_0}{\partial \alpha}+\frac{\delta S_{0
}}{\delta h^{\mu\nu}}P^{\mu \nu}
$$
\begin{equation}\label{diverges}
+(\kappa K_{\rho\sigma}-\overline{\eta}_\tau \overleftarrow{F}^\tau_{\rho\sigma})[\widetilde{s}(\widetilde{h}^{\rho\sigma})-s(\widetilde{h}^{\rho\sigma})]-\kappa L_\alpha[\widetilde{s}\widetilde{\eta}^\alpha-s\widetilde{\eta}^\alpha]\bigg].
\end{equation}
The last expression may be written as 
\begin{equation}\label{gorko}
\Gamma_1=S_{(0)gf}(\alpha+\delta \alpha, \beta+\delta\beta, \gamma+\delta \gamma,  \widetilde{h}^{\mu\nu}, \widetilde{\eta}^\alpha, K_{\mu\nu}, \overline{\eta}_\alpha, L_\sigma)-S^{rd}_{(0)gf}(\alpha, \beta, \gamma, h^{\mu\nu}, \eta^\alpha, K_{\mu\nu}, \overline{\eta}_\alpha, L_\sigma).
\end{equation}
Here  $S^{rd}_{(0)gf}$ is the gauge fixed action but with the terms $sh^{\mu\nu}$ and $s\eta^\alpha$ replaced by $s\widetilde{h}^{\mu\nu}$ and $s\widetilde{\eta}^\alpha$, that is, with the "non redefined" BRST variation of the "redefined fields". The important point is that not only  the gauge fixed action
$$
S_{(0)gf}=\int d^4x[L_s+(\kappa K_{\rho\sigma}-\overline{\eta}_\tau \overleftarrow{F}^\tau_{\rho\sigma})D^{\rho\sigma}_\alpha \eta_0^\alpha +\kappa^2 L_{0\alpha} \partial_\beta \eta_0^\alpha \eta_0^\beta],
$$
which can be written alternatively as
$$
S_{(0)gf}=\int d^4x[L_s+ (\kappa K_{\rho\sigma}-\overline{\eta}_\tau \overleftarrow{F}^\tau_{\rho\sigma})s h^{\rho\sigma} - L_{0\alpha} s\eta^\alpha].
$$
is BRST invariant,  but any expression of the form
\begin{equation}\label{mashine}
S^{rd}_{(0)gf}=\int d^4x[L_s+ (\kappa K_{\rho\sigma}-\overline{\eta}_\tau \overleftarrow{F}^\tau_{\rho\sigma})s(h^{\rho\sigma}+P^{\rho\sigma}) - L_{0\alpha} s(\eta^\alpha-Q_\beta^\alpha \eta^\beta)],
\end{equation}
has this invariant property as well. The BRST invariance of the last action follows from the fact that $s^2=0$, by an argument similar to \eqref{ufaa}  for Yang-Mills theories. It is in fact,  not difficult to see that this argument applies to the present context. Alternatively, one may check this BRST invariance directly by calculating
$$
s^2(h^{\mu\nu}+P^{\mu\nu})=s(sh^{\mu\nu}+\frac{\partial P^{\mu\nu}}{\partial h^{\alpha\beta}}sh^{\alpha\beta})=\frac{\partial P^{\mu\nu}}{\partial h^{\alpha\beta}\partial h^{\gamma\delta}}sh^{\gamma\delta}sh^{\alpha\beta}=0,
$$
the vanishing of this expression is justified since  $sh^{\mu\nu}=\kappa D_\epsilon^{\mu\nu}\eta^\epsilon$ and, due to the anti commuting nature of $\eta^\alpha$, the last expression is a product of a symmetric by anti-symmetric quantity, which is therefore zero. Furthermore the BRST action symmetry on the ghost is
$$
 \delta (\eta^\alpha-Q_\beta^\alpha \eta^\beta)=\delta\eta^\alpha -\bigg(\frac{\partial Q^\alpha_\beta}{\partial  h^{\mu\nu}}\delta h^{\mu\nu}\eta^\beta+Q_\beta^\alpha \delta \eta^\beta\bigg),
$$
from where
$$
s(\eta^\alpha-Q_\beta^\alpha \eta^\beta)=s\eta^\alpha+\frac{\partial Q^\alpha_\beta}{\partial  h^{\mu\nu}}sh^{\mu\nu}\eta^\beta-Q_\beta^\alpha s\eta^\beta.
$$
In deducing the last expression, the fact that the parameter $\epsilon$ present in the expression  $\delta$ anti commutes with $\eta^\beta$ was taken into account. Therefore
$$
\delta s(\eta^\alpha-Q_\beta^\alpha \eta^\beta)=\delta\bigg(\frac{\partial Q^\alpha_\beta}{\partial  h^{\mu\nu}}sh^{\mu\nu}\eta^\beta-Q_\beta^\alpha s\eta^\beta\bigg)
$$
$$
=\frac{\partial Q^\alpha_\beta}{\partial  h^{\mu\nu}}sh^{\mu\nu}s \eta^\beta \epsilon-\frac{\partial Q^\alpha_\beta}{\partial  h^{\mu\nu}}s h^{\mu\nu}\epsilon s\eta^\beta=0,
$$
as the expression for $s\eta^\alpha=-\kappa^2 \partial_\beta \eta^\alpha \eta^\beta$ contains two ghost fields and then  $\epsilon s\eta^\alpha=s\eta^\alpha \epsilon$. 
The last formula immediately leads to
$$
s^2(\eta^\alpha-Q_\beta^\alpha \eta^\beta)=0.
$$
This is the desired conclusion.

The reasoning above means that the currents in the action \eqref{mashine} are coupled to BRST invariant terms.
In fact it may  be concluded that any action of the form
\begin{equation}\label{difundir}
S^{gen}_{(0)gf}=\int d^4x[L_s+ (\kappa K_{\rho\sigma}-\overline{\eta}_\tau \overleftarrow{F}^\tau_{\rho\sigma})sF^{\rho\sigma}(h^{\alpha\beta}, \eta^\gamma) - L_{0\rho} sF^{\rho}(h^{\alpha\beta}, \eta^\gamma)],
\end{equation}
is a BRST invariant action as well. The new feature is that the currents $(\kappa K_{\rho\sigma}-\overline{\eta}_\tau \overleftarrow{F}^\tau_{\rho\sigma})$ and $L_\alpha$ are coupled to non standard BRST invariant terms, corresponding  redefinitions $F^{\rho\sigma}(h^{\alpha\beta}, \eta^\gamma)$ and $F^{\rho}(h^{\alpha\beta}, \eta^\gamma)$. 

Having concluded that the action $S^{rd}_{(0)gf}$ defined in \eqref{mashine} is BRST invariant,  it is still not concluded that the functional $\Gamma_1$ defined in \eqref{gorko} is BRST invariant. What it can be said is that it is represented as the difference between two BRST invariant functionals. This imitates the Yang-Mills case, however there is a big difference with that case since her one functional is invariant under the BRST action $s$ while the other is invariant with respect to $\widetilde{s}$ defined in \eqref{brstgrav2}. The task is to elucidate if $\Gamma_1$ is approximately BRST invariant, taking into account this property.

At present order in $\hbar$,  the formula \eqref{gorko} may be written as
$$
\Gamma_1=S_{(0)gf}(\delta \alpha, \delta\beta, \delta \gamma,  P^{\mu\nu}, Q^\alpha_\beta \eta^\beta, K_{\mu\nu},\overline{\eta}^\alpha, L_\sigma)
$$
\begin{equation}\label{gorkos}
-\int d^4x[ (\kappa K_{\rho\sigma}-\overline{\eta}_\tau \overleftarrow{F}^\tau_{\rho\sigma})s P^{\rho\sigma} - L_{0\alpha} sQ^\alpha_\beta \eta^\beta]+O(\hbar^2).
\end{equation}
Note that this action contains only first order terms in $\hbar$.  As for the Yang-Mills case, the previous action can be separated into a finite and divergent part by decomposing $$P^{\mu\nu}=P_f^{\mu\nu}+P_d^{\mu\nu},$$ and doing the same for $Q^\alpha_\beta$ and the parameters $\alpha$, $\beta$ and $\gamma$ of the gravitational part of the lagrangian. Following the procedure outlined in \eqref{aimitar} for Yang-Mills fields, redefine
\begin{equation}\label{aimitarse}
S_{0gf}\to S_{1gf}=S_{(0)gf}-\hbar \Gamma_1^d(S_{(0)gf}).
\end{equation}
This redefinition removes the divergent part in $\Gamma=S_{(0)gf}+\hbar \Gamma_1$.  The functional will be approximately BRST invariant up to $\hbar^2$, the argument is the same as the one outlined in  \eqref{disco}-\eqref{discu} for Yang-Mills theories. In addition, at first order in $\hbar$, it is clear that
$$
S_{1gf}=S_{(0)gf}(\alpha-\delta \alpha_d, \beta-\delta\beta_d, \gamma-\delta \gamma_d,  h^{\mu\nu}-P_d^{\mu\nu}, \eta^\alpha+Q^\alpha_{d\beta} \eta^\beta, K_{\mu\nu},\overline{\eta}^\alpha, L_\sigma)
$$
\begin{equation}\label{gorkoso}
+\int d^4x[ (\kappa K_{\rho\sigma}-\overline{\eta}_\tau \overleftarrow{F}^\tau_{\rho\sigma})s P_d^{\rho\sigma} - L_{0\alpha} sQ^\alpha_{d\beta} \eta^\beta].
\end{equation}
These considerations and the application of the method described in \eqref{reaching2}-\eqref{renormal2} lead to an expression for the renormalized effective action $\Gamma_r$ as an action with redefined fields
$$
\Gamma_r=S_{1gf}+ \hbar \Gamma^d_1+\hbar\Gamma_1^f=S_{(0)gf}+\hbar\Gamma_1^f
$$
$$
=S_{(0)gf}(\alpha, \beta, \gamma,  h^{\mu\nu}, \eta^\alpha, K_{\mu\nu},\overline{\eta}^\alpha, L_\sigma)
+S_{(0)gf}(\delta \alpha_f, \delta\beta_f, \delta \gamma_f,  P_f^{\mu\nu}, Q^\alpha_{f\beta} \eta^\beta, K_{\mu\nu},\overline{\eta}^\alpha, L_\sigma)
$$
$$
+\int d^4x[ (\kappa K_{\rho\sigma}-\overline{\eta}_\tau \overleftarrow{F}^\tau_{\rho\sigma})s P_f^{\rho\sigma} - L_{0\alpha} sQ^\alpha_{f\beta} \eta^\beta].
$$
At order $\hbar$ this can be written as
$$
\Gamma_r=S_{(0)gf}(\alpha+\delta \alpha_f, \beta+\delta\beta_f, \gamma+\delta\gamma_f,  h^{\mu\nu}+P^{\mu\nu}_f, \eta^\alpha+Q^\alpha_{f\beta} \eta^\beta, K_{\mu\nu},\overline{\eta}^\alpha, L_\sigma)
$$
\begin{equation}\label{gorkosol}
-\int d^4x[ (\kappa K_{\rho\sigma}-\overline{\eta}_\tau \overleftarrow{F}^\tau_{\rho\sigma})s P_f^{\rho\sigma} - L_{0\alpha} sQ^\alpha_{f\beta}\eta^\beta].
\end{equation}
The first term is the original action with the redefined fields, a feature that imitates the Yang-Mills situation \eqref{renormal}. The last two terms are non standard BRST invariant terms coupled to the currents, a feature that is different in Yang-Mills theories. 

Furthermore, at the linear order in $\hbar$ one has that  $P_f^{\mu\nu}(h^{\mu\nu})\sim P_f^{\mu\nu}(h^{\mu\nu}+P^{\mu\nu}_f)$ as the error is of second order. Also $Q_{f\beta}^\alpha(\eta^\gamma)\sim Q_{f\beta}^\alpha(\eta^\gamma+Q^\gamma_{f\delta} \eta^\delta)$.  In addition $s P_f^{\mu\nu}\sim \widetilde{s} P_f^{\mu\nu}$ and $sQ^\alpha_{f\beta}\eta^\beta\sim \widetilde{s}Q^\alpha_{f\beta}\eta^\beta$ as the difference $\widetilde{s}-s$
is of first order in $\hbar$ and  $P^{\mu\nu}$ and $Q^\alpha_{f\beta}\eta^\beta$ are also of that order, thus the difference affect only  orders higher than linear.  Therefore the last action can be written as 
$$
\Gamma_r=S_{(0)gf}(\alpha+\delta \alpha_f, \beta+\delta\beta_f, \gamma+\delta\gamma_f,  h^{\mu\nu}+P^{\mu\nu}_f, \eta^\alpha+Q^\alpha_{f\beta} \eta^\beta, K_{\mu\nu},\overline{\eta}^\alpha, L_\sigma)
$$
\begin{equation}\label{gorkosas}
-\int d^4x[ (\kappa K_{\rho\sigma}-\overline{\eta}_\tau \overleftarrow{F}^\tau_{\rho\sigma})\widetilde{s} P_f^{\rho\sigma} - L_{0\alpha} \widetilde{s}Q^\alpha_{f\beta}\eta^\beta]+O(\hbar^2).
\end{equation}
This action is written purely in terms of the first order redefined fields. In addition, it is BRST invariant at first order, since, as discussed above, the addition of the last two terms does not spoil BRST invariance due to the property $\widetilde{s}^2=0$. Furthermore $S_{(0)gf}$ is BRST invariant and the  last effective action solves the Slavnov-Taylor and ghost identities by construction, up to violating terms that are of order $\hbar^2$ or higher. This is due to the fact it is designed as a solution of \eqref{sumoso}, that is, the Slavnov-Taylor equation at first order. 

The above procedure can be repeated without obstructions to second order.   By taking into account the approximately BRST invariant  $\Gamma_1^r$  defined in \eqref{gorkosas} as initial action, which solves the desired equations at the given order,  by them repeating all the above procedure exactly, the second order effective action $\Gamma_2^r$ can be constructed, which will be again BRST invariant up to order $\hbar^3$. This action will be given again in terms of $S_{(0)gf}$ with redefined fields,  with the possible addition of non standard BRST terms coupled to the currents, analogous to the last two terms in \eqref{gorkosas}. This procedure has no obstruction to be repeated inductively at $n$-th order, thereby leading to the renormalized action 
$$
\Gamma_n^r=S_{(0)gf}^{rd}(\alpha+\delta_n \alpha_f, \beta+\delta_n\beta_f, \gamma+\delta_n\gamma_f,  h^{\mu\nu}+P^{\mu\nu}_{nf}, \eta^\alpha+Q^\alpha_{nf\beta} \eta^\beta, K_{\mu\nu},\overline{\eta}^\alpha, L_\sigma)+O(\hbar^{n+1}),
$$
where the subindex $n$ indicates that the field has been redefined $n$-times. Furthermore the index $rd$ indicates  that the action may contain several new BRST invariant terms coupled to the currents $ (\kappa K_{\rho\sigma}-\overline{\eta}_\tau \overleftarrow{F}^\tau_{\rho\sigma})$  and $L_\alpha$ due to those field redefinitions. The key point is that order by order the effective action is approximately BRST invariant, which is the main purpose of the present exposition. It is interesting that the methods of multiplicative renormalization in Yang-Mills theories were useful in the present context, even though the renormalizability for quadratic gravity theories is not multiplicative. 

 \section{Discussion}
 In the present text, several important features about multiplicative renormalization in Yang-Mills theories and the renormalization of quadratic gravity were discussed, exploiting the analogies between both scenarios as much as possible.  A key point for interpreting the counter terms appearing in the gravity case \cite{stelle2} is that the effective action is expressed, order by order in $\hbar$,  as the original classical action in terms of redefined fields, together with non standard though BRST invariant coupling to the ghost currents. These couplings are absent in Yang-Mills theories, nevertheless they do not spoil BRST symmetry at a given order. 

 The discussion given along the text about renormalizabilty of quadratic gravity of course,  does not cure the problem of unitarity of these theories.   A possible and interesting investigation is to study how this procedure may be generalized to the models described in  \cite{mannheim1}-\cite{donogue}. In particular the formulas (4.35)-(4.45) of reference \cite{salve} present a path integral method for dealing with theories such as the present, which are of higher order  and contain apparent instabilities. This method involves an additional factor in the measure of the path integral, indicated by $C$ in that reference, which involves the metric determinant and the volume of the space slices. This factor does not affect perturbative calculation around flat spaces, as shown in \cite{popiv}, since they involve a term proportional to $\delta^4(0)$ which is cured by dimensional regularization. For perturbation around curved spaces, this determinant may play a role.

 Finally, an interesting lead may be to generalize the cohomological methods developed for Yang-Mills theories \cite{kluberg}-\cite{joglekar} to the gravitational case. Perhaps the two weak points in the discussion about the renormalizability of quadratic gauge theories is the assumption of the $\hbar$ expansion for the effective action $\Gamma$  and the validity for the cohomological method for gravity theories. The formal lines developed in \cite{henneaux}-\cite{anselmi3} may be useful for understanding this second point. In any case, both assumptions deserve to be investigated more closely.
 
 \section*{Acknowledgements}
O. S is supported by CONICET, Argentina and the Grant
PICT 2020-02181.

 \end{document}